\documentstyle[aps,epsfig,subfigure]{revtex}

\begin{document}

\title{Active optical fibres in modern particle physics experiments}

\author{Carsten Patrick Achenbach\thanks{
     Present address: Institut f{\"u}r 
     Kernphysik, Joh.\ Gutenberg-Universit{\"a}t Mainz, 
     J J Becher-Weg 45, 55099 Mainz, Germany.
     Tel.: +49--6131--3925831; fax: +49--6131--3922964.
     {\it E-mail:}~patrick@kph.uni-mainz.de}}
\address{University of Oxford, Sub-department of Particle Physics, 
	Denys Wilkinson Bld., Keble Rd., Oxford, OX1 3RH, UK}
\date{27 July 2003}

\maketitle

\begin{abstract}
  In modern particle physics experiments wavelength-shifting and
  scintillating fibres based on plastic polymers are used for tracking
  and calorimetry. In this review the role of photon trapping
  efficiencies, transmission functions and signal response times for
  common multimode active fibres is discussed. Numerical simulations
  involving three dimensional tracking of skew rays through curved
  fibres demonstrate the characteristics of trapped light. Of
  practical interest are the parametrisations of transmission
  functions and the minimum permissible radius of curvature. These are
  of great importance in today's experiments where high count rates
  and small numbers of photoelectrons are encountered. Special
  emphasis has been placed on the timing resolution of fibre detectors
  and its limitation due to variations in the path length of generated
  photons.
\end{abstract}

\section{Introduction}

Optical fibres with large core diameters, i.e.\ where the wavelength
of the light being transmitted is much smaller than the fibre
diameter, are commercially available and readily fabricated, have good
timing properties and allow a multitude of different
designs. Multimode fibres are useful for short data-bus connections,
local area networks and for multiplexing and sensor technologies.
Since their first appearance in charged particle detectors of the
early 1980s active optical fibres have also played an important part
in the field of nuclear and particle physics. Optical fibres are
commonly produced from glass, plastic and synthetic fused silica,
often called silica or quartz fibre. Each type has its own advantages
and drawbacks. Early glass materials were based on Cerium (Ce$^{3+}$
oxide) and have attracted some attention. For charged particle
detection, plastic scintillator compositions have been emerged as the
far superior material. In contrast, for data communications
applications, silica fibre is the overwhelming choice. Nowadays, the
low costs of plastic base materials make it possible for many particle
physics experiments to use plastic fibres in large quantities.

Light is generated inside an active fibre either through interaction
with the incident radiation (scintillating fibres) or through
absorption and re-emission of primary light (wavelength-shifting
fibres). A small fraction of the emitted light is guided via total
internal reflection to the fibre end where it is detected by
visible-light photon sensors. The great interest in fibres is based on
the fast signal response of organic scintillators and the high spatial
precision and easy handling of fibres. As the output signals of many
photon sensors are short on the time-scale of electronic circuits and
data acquisition systems, these detectors can be operated at high
count rates. In case the active fibres are located inside a strong
magnetic field region, the fibres can be spliced on clear fibres that
have a higher transmission, so that the photon sensors can be placed
outside the field region. This is also done if the active area of a
scintillating fibre detector has to be restricted to minimise
background count rates.

In response to the need for precise and fast detectors Borenstein
reported on the properties of plastic scintillating fibres in the year
1981. His group had measured decay constants of a few nanoseconds in
fibres.  In these years, development work was focused on vertex
detectors for fixed target experiments. In addition, some experiments
used scintillating fibres as active targets for the study of rare
processes. Active targets were formed from coherent arrays of
scintillating glass fibres, plastic fibres, or capillaries filled with
liquid scintillators and the targets were viewed by image
intensifiers. Throughout the 1980s and 1990s substantial efforts have
been devoted to the development of better glass and plastic
scintillation materials and the detection of ionising radiation with
scintillating fibres has been generally practised in many different
ways. The main applications of active fibres in the present generation
of particle physics experiments are large-area tracking detectors and
fine-sampled calorimeters.

Trackers are sub-detectors which surround the interaction point to
reconstruct charged particle tracks and their vertex positions at
lepton or hadron accelerators. A modern system can be found at the
Fermi National Accelerator Laboratory where the D$\emptyset$ Central
Fiber Tracker comprises 71,680 multi-cladding fibre
channels~\cite{DZERO1995}. One of the most exhaustive of all fibre
systems is the CHORUS tracker~\cite{CHORUS1998}. It is based on a
total of about 1.2 million plastic scintillating fibres of 0.5\,mm
diameter which are read-out via an opto-electronic system comprising
image intensifier tubes and CCD cameras (Charged Coupled Devices) in
series. The long baseline neutrino oscillation experiment K2K also
employs, as a component of its near detector, a scintillating fibre
tracker with 274,080 fibres in total~\cite{K2K2000}. For its signal
processing groups of 11,420 fibres are glued together to make single
bundles of 12\,cm diameter each with an opto-electronic read-out
system similar to the CHORUS one. For the construction of the highly
segmented tracker of the ATLAS experiment more than 600,000
wavelength-shifting fibres have been used~\cite{ATLAS1994}. For fibre
trackers the basic element is a fibre doublet ribbon, which is formed
from two single layers of fibres, with one of the fibre layers set off
relative to the other by half a fibre spacing. The stacking of fibres
provides a higher light yield per channel and enough spatial overlap
to avoid relying on the detection of events with only a grazing
contact with the charged particle.  A charged particle which crosses
the gap between two fibres in one of the layers is likely to traverse
the full fibre diameter in the other layer. It has been shown that a
high detection efficiency and a good spatial resolution can be
achieved with multiple doublets.  The good timing resolution of such
tubes enables a higher level event-selection in the early stage of the
trigger of large detector systems. Used in conjunction with other
detectors like calorimeters or muon spectrometers, such trigger can
provide a powerful signature for identification of electrons, muons,
photons and event vertices. Accordingly, multi-layered structures of
stacked scintillating fibres coupled to multi-anode photomultipliers
became the preferred choice for some fast trigger detectors, the
COMPASS trigger~\cite{Horikawa1999} being one recent example.

Fibre calorimeters consist of dense absorber materials sampled with
scintillating fibre planes to achieve a very compact geometry. Fibre
calorimeters are found, for example, in the muon (g-2)
experiment~\cite{Sedykh2000} at Brookhaven and in the KLOE
detector~\cite{KLOE1996} at the DA$\Phi$NE accelerator of the INFN
LNF, where lead foils are interleaved with layers of scintillating
fibres. In some cases, bulk scintillators are read out via
wavelength-shifting fibres, most recently in the MINOS
experiment~\cite{MINOS1998} where fibres are embedded in 8\,m long
scintillating bars. The approved experiments for the LHC collider at
CERN are relying heavily on active fibres, too. The LHCb experiment
uses 6000 detector cells with wavelength-shifting fibre
read-out~\cite{LHCb1998} and the CMS experiment uses
wavelength-shifting fibres embedded in scintillator plates for its
sampling hadronic calorimeter~\cite{CMS1997}. Fibre calorimeters can
achieve energy resolutions comparable to full absorption calorimeters.

At high luminosity accelerators a new generation of magnetic
spectrometers is being developed for fixed target experiments to
reveal the nucleon structure with electromagnetic probes. Up to now
large magnetic spectrometers have mainly used various types of wire
chambers as coordinate detectors for momentum determination in the
dispersive plane. But these detectors are slow and in the high
particle flux environments of the high luminosity accelerators very
fast detectors are indispensable.  Furthermore, active elements of low
density with long radiation and interaction lengths are necessary to
avoid multiple scattering and energy loss of particles. For the same
reason, mounting structures and other inactive materials have to be
minimised within the detector system's volume. These requirements can
be met by stacks of scintillating fibres which are a good compromise
between signal to noise ratio and trigger or tracking efficiency on
the one side and the amount of material traversed by the particles on
the other side. Alternative detector concepts include the newly
developed carbon plated kapton straw tubes with minimised drift times.
In addition, there has been an impressive progress during the last
couple of years on gaseous micro-pattern detectors. These detectors
have much higher rate capabilities than wire chambers. In a detector
called Micromegas the gas amplification happens between a metal mesh
several $\mu$m thick and a printed circuit board with metal strips
about 0.2\,mm apart~\cite{Giomataris1996}. Another detector called Gas
Electron Multiplier (GEM), first introduced in November 1996 by
Sauli~\cite{Sauli1997}, is very promising. Its key element is the GEM
electrode, consisting of a polymer film about 50\,$\mu$m thick,
metallised on both sides, with a regular array of small holes. For a
spacing of 140\,$\mu$m the holes have diameters of about
70\,$\mu$m. In stacks of several GEMs amplification and induction gaps
are separated and very high gas amplifications can be reached. Another
type of detector to be considered as competitive to conventional
detector concepts in future fixed target experiments are silicon
micro-strip or micro-pad detectors. These are the detectors of choice
for the LHC experiments ATLAS and CMS. They provide an extremely good
spatial resolution but their timing resolution is limited to $\sigma
\le 100\,$ns by the signal to noise ratio of the solid state
element. Another disadvantage of such detectors is their relatively
large thickness. From the above discussion one may conclude that the
fastest coordinate detectors presently available---scintillating
fibres---are playing a decisive role for the latest experiments in
hadron physics where luminosities and particle fluxes are challenging.

Thus, the interest in active fibres by experimental particle
physicists is large. In recent years, additional applications have
emerged in medical and biological dosimetry, in electron or ion beam
monitoring, in activity studies of radioactive waste, and in X-ray and
synchrotron radiation detection. In contrast, the basic theory
describing the propagation of photons in fibres is not commonly known
to particle physicists. The strict treatment of small diameter optical
fibres involves electromagnetic theory applied to dielectric
waveguides, which was first achieved by Snitzer~\cite{Snitzer1961} and
Kapany~\cite{Kapany1963}. This is not a simple undertaking, especially
for bent fibres where an eigenvalue equation is not available. To
solve this problem many approximation techniques have been developed
which are reasonably accurate for single-mode fibres. In multimode
fibres, however, the electromagnetic fields get substantially modified
by any curvature~\cite{Winkler1979}. Light losses in small diameter
fibres with uniform curvature have been calculated numerically but the
method becomes extremely difficult for fibres of large
diameters. Although these approaches provide insight into the
phenomenon of total internal reflection and eventually lead to results
for the field distributions and the radiation from curved fibres, it
is advantageous to use ray optics for large diameter fibres where the
waveguide analysis is an unnecessary complication. The optics of
meridional rays in fibres was developed in the 1950s~\cite{Kapany1957}
and can be found in numerous textbooks,
e.g.~\cite{Kapany1967,Allan1973,Ghatak1998}. Since then, the
scientific and technological progress in the field of fibre optics has
been enormous.

This paper is organised as follows: section~2 starts with a review of
active optical fibres in the field of detection and measurement of
ionising radiation and charged particles. In section~3 a simulation
code is outlined that performs a three-dimensional tracking of light
rays in cylindrical fibres. The first part of section~4 discusses the
trapping of light rays and the quantitative distinction between skew
and meridional rays. Then, it continues with a comprehensive overview
on the light yield of active fibres where special emphasis is placed
on light losses in sharply curved fibres. The chapter ends with a
paragraph on the radiation resistance of plastic fibres. The timing
resolution of active fibres is an important and prevailing issue that
is presented in section~5. Finally, the conclusions review the current
research on active fibres for modern particle physics
experiments. References on latest developments in scintillating fibre
technology are given throughout the text as they apply to
charged-particle detection.

\section{General discussion}

\begin{figure}[htbp]
  \begin{center}
    \epsfig{width= 0.5 \textwidth, file= 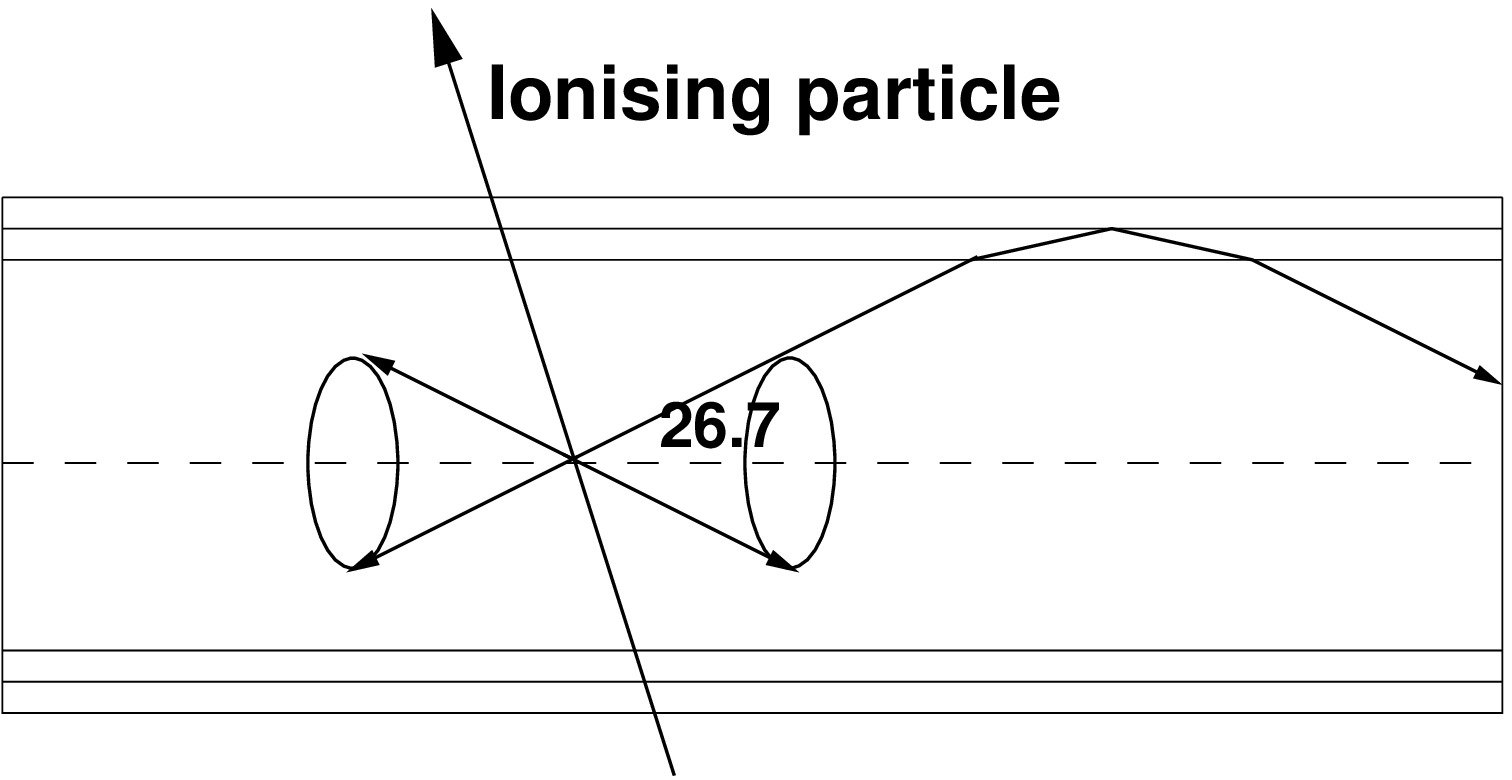}\\[5mm]
    \caption{Cross sectional view of a double cladding fibre. The 
	cone of trapped light is limited in the meridional 
	approximation by a critical axial angle $\theta_{\it crit} 
	= \arccos n_{\it clad'}/n_{\it core}$, that equals 
	26.7$^\circ$ for a ``standard'' fibre.}
    \label{fig:sketch}
  \end{center}
\end{figure}

For charged particle detection mostly fibres produced from plastic
polymers are used. For several reasons they are better suited than
alternative candidates. For example, they have intrinsically higher
efficiency, faster signal response times, and lower material
densities. In general, a typical fibre consists of a core coated with
a thin ($2-5\,\mu$m) transparent cladding with a smaller index of
refraction than the core has.  The plastic cores of active fibres
include several components. The base material ($x > 98\%$ by weight)
is an organic polymer, such as polystyrene (PS) or polyvinyltoluene
(PVT). It is doped with organic molecules, mostly aromatic compounds,
which emit scintillation\footnote{Scintillation is an example of
radioluminescence. Scintillators may be organic and inorganic solids,
liquids, and gases.} light. Total internal reflection at the
core-cladding interface allows for an efficient transport of light
over many metres. The cladding itself provides a protective layer
around the fibre core. A fibre with the refractive index constant over
the fibre core cross-section is called a {\it step index} fibre. The
most common type of fibre used in particle physics consists of a
polystyrene-based core of refractive index $n_{\it core}=$ 1.6 and a
thin polymethylmethacrylate (PMMA, C$_5$H$_8$O$_2$) cladding of
refractive index $n_{\it clad}=$ 1.49 (indices are given at a
wavelength of 590\,nm). Throughout this paper I will refer to this
formulation as ``standard''. A more recent cladding material is
fluorinated methylmethacrylate (MMA) with $n_{\it clad'}=$
1.42. Single MMA cladding fibres have shown a poor performance in
terms of absorption and mechanical stability, so that MMA now is only
used in double cladding fibres. These have an inner PMMA cladding and
an outer MMA cladding, which leads to a significantly increased
trapping efficiency and light yield. In figure~\ref{fig:sketch} a
cross section through a double cladding fibre is shown to illustrate
the cone of trapped light in the meridional approximation. For a
double cladding fibre the critical axial angle is given by
$\theta_{\it crit} = \arccos n_{\it clad'}/n_{\it core} = 26.7^\circ$.
In principle an active cladding is possible, but uncertain mechanical
compatibility issues between the core and a chemically modified
cladding have prevented such a development up to now. In this paper I
will consider round fibres only, since edged cross-sections tend to
entail in the manufacturing process serious problems of homogeneity at
the edges.  Typical diameters of fibres used in calorimetry are in the
range of 0.5 -- 1.5\,mm, for tracking detectors the diameter is of the
order of 10 -- 100\,$\mu$m. Obviously, the fibre dimensions together
with their geometrical overlap define the spatial
precision of a tracker. An early work on active plastic fibres can be found
in~\cite{White1988}. In the year 1995, Leitz~\cite{Leutz1995} wrote an
excellent review of scintillating fibres in particle physics,
including working conditions of fibres, the scintillation process and
photoelectron counting.

After the passage of a charged particle many of the atoms in the
medium will be excited into higher energy levels. The electronic
levels have a typical energy spacing of $\Delta E \simeq
4\,$eV. Transitions from the ground state to the excited state occur
with no change in interatomic distance. The excited molecule quickly
loses vibrational energy to arrive in an electronic state S$^*$. In
non-scintillating materials the excitation energy is given up in the
form of heat or lattice vibrations. In a scintillator, however, some
of the excitation energy is released as electromagnetic radiation.
These compounds contain bound benzene rings, e.g.\ p-terphenyl. The
electronic state S$^*$ of a scintillator decays to some vibrational
level of the ground state within a characteristic decay time, $\tau
\le 2\,$ns. The emitted light is not self-absorbed as the light
propagates along the fibre by a further S$_0 \rightarrow$ S$^*$
transition because of the differing shapes of the excited and ground
state energy levels as a function of interatomic spacing.  The net
result is a disparity between the absorption and the emission spectrum
of the molecule, where the details of both spectra are governed by the
electronic structure of the molecule. The difference between the
wavelengths of the peak positions is called Stokes' shift. It is the
excitation of the electrons within the $\pi$-orbitals that results in
the scintillation.

In polystyrene the absorption peak is far from the p-terphenyl
emission, although the base is opaque to this light because of
Rayleigh scattering. Usually, the base scintillators are not good
intrinsic light transmitters. The fluorescence yield, that is the
ratio of the number of excited atoms which emit a photon to the total
number of excited atoms, of pure polystyrene is rather poor, $Q\approx
0.03$. A shift of the emission light into the base material's
transparent region ($\lambda > 400\,$nm for polystyrene) can be
achieved by adding a fluorescent dopant to the monomers of the core
base material for combined polymerisation. Most modern fibres are
working with a one-component system (3HF or PBBD), where the
sufficiently high concentration (molar fraction $x \ge 1\%$) of the
dopant allows local, non-radiative F\"orster transitions from the
excitation site to the dopant. F\"orster transitions are fast ($\Delta
t < 1\,$ns) and resonant dipole-dipole transitions. Thereafter, the
excitation energy is stored in the excited dopant molecule. Usually,
the dopant is chosen to have a large Stokes' shift which leads to a
small self-absorption of the secondary light. The fluorescence yield
can be enhanced to values of $Q= 0.8-0.9$. Several families of
compounds with these properties (fast, efficient, and low
self-absorption) have been found. Among these are the hydroxyflavons,
hydroxybenzothiazoles and hydroxybenzoxazoles, from which 3HF
(3-hydroxyflavone) has been produced.

Some dopants like p-terphenyl or PBD need additional
wavelength-shifting components in very low concentrations, for example
1,4-di-(2-(5-phenyloxazolil))-benzene (C$_{24}$H$_{16}$N$_2$O$_2$),
called POPOP, or bPBD (C$_{18}$H$_{14}$). Their absorption bands
overlap the emission band of the primary dopant and the primary light
is absorbed by the wavelength shifter before it can get attenuated by
other processes. The self-absorption of the secondary light is smaller
than the self-absorption of the primary light because of the low
concentration (molar fraction $x < 1\%$) of the wavelength-shifting
molecules. The minimal self-absorption also helps to eliminate
cross-talk between neighbouring fibres and the long emission
wavelengths provide improved immunity to the effects of radiation
damage. More details of the scintillation process are discussed in
depth by Birks~\cite{Birks1964}.

Wavelength-shifting fibres are very similar in structure to the
scintillating fibres discussed earlier. A wavelength-shifting fibre
uses a non-scintillating base material doped with a fluorescent dye to
shift external light from a scintillator to longer wavelengths.  The
most common formulation is an absorption peak in the blue region of
the spectrum and an emission peak in the green or yellow region, where
a wide spectral overlap between the emission spectrum of the
scintillator and the absorption spectrum of the wavelength-shifting
fibre is necessary. The fibres may be positioned in grooves machined
into the surface of plastic scintillators bars. The quantum efficiency
of the wavelength-shifter is usually high, i.e.\ in the range of
$70-80$\%.

In the fibre core a certain fraction of the scintillation light,
called {\em core} light, is trapped by total internal reflections at
the core-cladding interface. Light not trapped in the fibre core is
refracted into the cladding and again some portion of this light is
trapped by total internal reflections at the cladding-air interface,
called {\em cladding} light.  Because of the stepping refractive
indices the trapping efficiency for cladding light is much larger than
the one for core light.  In general, the relative contribution of each
light component depends on the distance of the detector from the
emission point and the attenuation lengths. Most light detected after
short fibre lengths can be related to the cladding light, but the
cladding light is attenuated to a greater degree than the core
light. The losses of cladding light are mainly caused at the
cladding-air interface which is exposed to the environment and
therefore will degrade by abrasion or accumulation of impurities. The
removal of the cladding light can be useful and is accomplished by
coating the fibre with an extra-mural absorber. It is also known that
internal reflections being less than total give rise to so-called {\em
leaky} or non-guided modes, where part of the electromagnetic energy
is radiated away at the reflection points. An extra-mural absorber
eliminates those components together with most of the cladding light.
Sometimes a diffuse reflecting paint, usually TiO$_2$, is used for
gluing stacks of fibres together.  In these applications the
extra-mural absorber suppresses optical cross-talk between adjacent
fibres, which occurs when untrapped photons, i.e.\ about 90\% of all
light, are trapped in neighbouring fibres after being scattered. A
trapped ray may be categorised by its path along the fibre. The path
of a meridional ray is confined to a single plane, all other modes of
propagation are known as skew rays.  The projection of a skew ray on a
plane perpendicular to the fibre axis changes its orientation with
every reflection. In the special case of a cylindrical fibre all
meridional rays pass through the fibre axis.

The short and fast light pulses of active fibres are detected by
photo-effect based light detectors.  Several sensors for the detection
of the fibre light have been investigated. In most cases the detection
is performed by means of conventional photomultiplier tubes. Their
photocathodes are mainly composed of semi-conducting photo-emissive
materials (bialkali, trialkali and others) which are evaporated as
thin films on the inside of optic front windows. The quantum
efficiency of a typical photocathode is good ($Q.E. \approx 20-25\%$)
in the blue spectral region and fair to poor ($Q.E. \approx 5-10\%$)
in the yellow and green region. This combination of fast active
elements with modern read-out devices entails the characteristics of
low noise, large gain, good linearity and good timing resolution. The
details of the coupling depend strongly on transmission range and
refractive index of the window material and on the quantum efficiency
of the photocathode. The refractive index of front windows ranges from
1.4 (LiF) over 1.5 (crown glass, borosilicate, quartz) to 1.95
(YalO$_3$), the transmission edge ranges from 120 to 350\,nm.  Mineral
oils and optical couplants have been used successfully as optical
coupling media bridging the difference in refractive indices. For most
fibre to photomultiplier interfaces transmission values above 95\% are
readily achieved. Losses can occur because of longitudinal, lateral
and angular misalignment.

Position-sensitive photomultipliers are especially suitable for fibre
read-out because of the good match between photomultiplier
segmentation and common fibre diameters, offering an important
reduction in size and cabling with respect to conventional
tubes. Since the pioneering work of Kuroda~\cite{Kuroda1981} such
tubes have been developed in order to meet the demands on precise and
reliable tracking devices under high-rate
circumstances~\cite{FAROS1995,FAROS1996,Agoritsas1998}. In recent
years they have been continously improved. In modern experiments,
fibre bundles involving a rather large number of channels are easily
read out via position-sensitive photomultipliers, while the use of
single channel photomultipliers is no longer economical in terms of
cost and space requirements. Three different types of
position-sensitive photomultipliers are in use: a) multi-anode
photomultiplier with a grid-like dynode structure and segmented anode
pixels, (b) multi-dynode photomultiplier with a separate dynode
structure for each channel and (c) multi-channel photomultiplier where
each channel additionally has its own photocathode and glass
window. The two most important parameters to characterise
position-sensitive photomultipliers is the amount of cross-talk
between adjacent channels and the channel to channel gain variations.
The first generation of position-sensitive photomultiplier tubes
suffered from high cross-talk and large gain variations.

The Hamamatsu Photonics~\cite{Hamamatsu} R5900 series is a rather new
development which has been chosen by many groups, e.g.\ the MINOS
collaboration~\cite{MINOS1998}, as their fibre read-out device. The
performance of this tube was first reported in~\cite{Yoshizawa1997}
and many comprehensive tests have been conducted since then, see
e.g.~\cite{Enkelmann1998}. The drawbacks of early position-sensitive
photomultipliers have been greatly reduced; the tube exhibits very
little cross-talk and a high gain uniformity between channels. It
comes in a very compact design and six different anode geometries
suitable for fibre read-out, one particularly useful type of anode
shape is the M-16, in which the $17.5 \times 17.5$\,mm$^2$ bialkali
photocathode is divided into 16 pixels each with a sensitive area of
$4 \times 4$\,mm$^2$ per pixel. The cathode is followed by individual
metal channel dynodes incorporating 10 to 12 stages and the output
signal is read out from independent multiple anodes. Single fibres or
bundles of up to seven fibres can get coupled to one pixel. 

Photomultiplier tubes are limited in their intrinsic energy resolution
by fluctuations in the number of secondary electrons produced at the
first dynode. This limitation can be overcome by using hybrid
photomultiplier tubes with their excellent multiple photon separation
and high efficiency. A hybrid photomultiplier tube consists of a
reversely biased silicon P-I-N diode, in which highly accelerated
photoelectrons create a few thousand electron-hole pairs with much
smaller statistical fluctuations. The quantum efficiency of silicon
photodiodes can be as high as 70\% for visible wavelengths. In special
cases, image intensifiers with CCD cameras, solid state
photomultipliers or so-called Visible Light Photon Counters (VLPC)
have been employed in active fibre read-out. The VLPC has been
developed by Rockwell Int.\ Science Center and is used e.g.\ in
experiment E835 at the Fermi National Accelerator
Laboratory~\cite{Ambrogiani1998}. It can reach 85\% quantum
efficiency. Some studies on silicon based avalanche photodiodes have
been performed~\cite{Baer2000} and it was demonstrated that it is
possible to use this type of photodiodes for room-temperature fibre
read-out, too. The diodes consist of a compact semiconductor element
operated with at reverse bias voltage near the breakdown voltage. The
basic structures include an absorption region and a multiplication
region with a high electric field. The multiplication region is broad
enough to provide a useful gain of at least 100 by impact
ionisation. Recently, there has been some progress in improving gain
and stability and first arrays of avalanche photodiodes have become
available to cope with a large number of fibres closely lined up.

The signal amplitude from a fibre detector, quantified by the number
of detected photoelectrons, is a complex factor involving the light
output of the fibre, the light collection efficiency, the coupling of
the fibre to the photon sensor, and the sensor's characteristics. In
many applications the number should be as large as possible to
discriminate the signal against dark pulses and electronic noise. It
is instructive to work through a numerical estimate.  The following
example gives a parametric factorisation of the signal amplitude for a
scintillator bar with a wavelength-shifting fibre of approximately
3\,m length, read out with a multi-anode photomultiplier. The
scintillator bar is exposed to a uniform illumination of charged
particles from one direction at normal incidence at the far end of the
scintillator. The source of the signal is the luminescence of the
scintillator, numerically equal to the number of scintillation photons
produced at the site of ionising radiation. It can be estimated by:
\begin{eqnarray}
  {\mathcal L}_{scin} & = & \Delta E \times \frac{d N_{scin}}{d E}\\ &
	= & \frac{d E}{d\rho x}\Big|_{mip} \times \rho x
 	  \frac{d N_{scin}}{d E} \simeq 35,000 \,
\end{eqnarray}
where a mean energy deposition corresponding to minimum ionising
particles $-dE/d\rho x|_{mip}= 1.68$\,MeV$/$cm$^2$, the density of
plastic scintillator $\rho= 1.032$\,g$/$cm$^3$, the average path
length of a charged particle traversing the scintillator of height $x=
2$\,cm, and an absolute light yield, $d N_{scin}/d E$, of 10,000
photons produced per MeV of deposited energy in polystyrene is
assumed.  The light yield corresponds to $\sim$1 excitation per
4.8\,eV and a quantum efficiency $Q\sim 4-5\%$ of the scintillator.

The light collection and transfer factor, $\epsilon_{coll}$, of a
scintillator bar is of the order of 50\%. It is dependent on the
geometry and the effective attenuation length of the scintillator,
$\Lambda$, the reflectivity of the scintillator surface, $R$, and the
number of fibres per scintillator bar, $i$. For complicated geometries
the exact number may be evaluated by Monte Carlo simulations. In case
of a simple geometry it can be estimated. In infinite long bars the
probability of a photon to hit a fibre of diameter $2\rho$ can be
written as $p_i= i 2\rho/l$ where $l$ is the distance between
adjacent fibres, in this example identical to the width of the bar,
$w$. The probability of a photon being absorbed in the scintillator is
simply $p_\Lambda = 1 - e^{-P/\Lambda} \approx l/(\Lambda\,
\cos\theta)$, where $P$ is the path length of the photon between two
fibres. The probability for the photon not to be reflected can be
written as $p_R = 1 - R$. This leads to a light collection efficiency
of
\begin{equation}
  \epsilon_i = \frac{p_i}{l/(\Lambda\, \cos\theta) + (1 - R) + p_i}\ ,
\end{equation}
In this model the increase in light yield when using two fibres
instead of one is equal to $\epsilon_2 / \epsilon_1 = 57\%/40\% =
1.43$. The simple model has been verified to be sufficiently accurate
by using a ray tracing code to find the fraction of scintillation
photons which are absorbed in a wavelength-shifting fibre located in a
groove in a rectangular scintillator.

Photons are emitted isotropically within the fibre. Using the
meridional approximation the trapping efficiency for core light
emitted in one axial direction of the fibre is $3-5\%$ for a single
cladding fibre. For this example the transmission functions of fibre
and photomultiplier entrance window are assumed to be $T_{fibre}\simeq
70\%$ and $T_{PMT}\simeq 97\%$. The emission spectrum is a
characteristic of the scintillation material and the spectral quantum
efficiency of the detector is a characteristic of the photocathode. An
wavelength averaged value of $Q.E.\simeq 16\%$ is reasonable. Finally,
the number of photoelectrons in a typical fibre experiment can be
estimated by combining the above values:
\begin{equation}
  N_{p.e.} = {\mathcal L}_{scin} \times \epsilon_{coll} \times \Omega_{1/2}
	\times T_{fibre} \times T_{PMT} \times Q.E. \approx 25\ .
\end{equation}
Hence, a typical detected photoelectron yield is of the order of
$\bar{N} \approx 20 - 30$. This value is consistent with measurements.

\section{Three-dimensional Tracking of Light Rays}

To illustrate the characteristics of trapped light I use results from
a programme which simulates emission and propagation of light
rays. The motivation for writing the programme was to understand the
loss of light in sharply curved fibres. Since the analytic analysis of
the passage of skew rays along a curved fibre is exceedingly complex,
a Monte Carlo technique had to be applied. This type of numerical
integration using random numbers is a standard method in the field of
particle physics and is now practical given the CPU power currently
available. Detailed results of this programme on the light acceptance
and propagation in straight and curved multimode active fibres can be
found in~\cite{Achenbach2003}.

A light ray is followed by the simulation until it is absorbed or
detected. Quantities to be delivered by the programme are the
proportion of light detected, and the arrival time distribution, or
the various ways light rays may be lost. On its path the ray is
subject to attenuation, parameterised firstly by an effective
absorption coefficient and secondly by a reflection coefficient. At
the core-cladding interface the ray can be reflected totally or
partially internally. In the latter case a random number is compared
to the reflection probability to select reflected rays.

Light rays are randomly generated on the circular or rectangular
cross-section of a fibre with radius $\rho$ or width $w$ and height
$h$. An arbitrary ray is defined by its axial and azimuthal or skew
angle. An advantage of this method is that any distribution of light
rays can easily be generated. The axis of the fibre is defined by a
curve $z= f(s)$ where $s$ is the arc length. For $s < 0$, it is a
straight fibre along the negative $z$-axis and for $0 < s < L_F$, the
fibre is curved in the $xz$-plane with a radius of curvature $R_{\it
curv}$. In particular, the curve $f(s)$ is tangential to the $z$-axis
at $s = 0$.

Light rays are represented as lines and determined by two points,
$\vec{r}$ and $\vec{r}^{\,\prime}$. The points of incidence of rays
with the core-cladding interface are determined by solving the
appropriate systems of algebraic equations. In the case of a straight
fibre the geometrical representation of a straight cylinder or box is
used resulting in a quadratic equation. Its positive solution defines
the point of incidence, $\vec{r}_R$, on the fibre wall. In the case of
a cylindrical fibre curved in a circular path, the cylinder equation is
generalised by a torus equation. The coefficients of this fourth
degree polynomial are real and depend only on $R_{\it curv}$ and the
vector components of $\vec{r}$ and $\vec{r}^{\,\prime}$ up to the
fourth power. In most cases there are two real roots, one for the
core-cladding intersection in the forward direction and one at
$\vec{r}$ if the initial point already lies on the cylinder wall. The
roots are found using Laguerre's method~\cite{Recipes1992}. It
requires complex arithmetic, even while converging to real roots, and
an estimate for the root to be found. The routine implements a
stopping criterion in case of non-convergence because of round-off
errors. The initial estimate is given by the intersection point of the
light ray and a straight cylinder that has been rotated and translated
to the previous reflection point. A driver routine is used to apply
Laguerre's method to all four roots and to perform the deflation of
the remaining polynomial. Finally the roots are sorted by their real
part. The smallest positive, real solution for $m$ is then used to
determine the reflection point, $\vec{r}_R$.

After the point of incidence has been found, the reflection length and
absorption probability can be calculated. The angle of incidence,
$\alpha$, is given by $\cos{\alpha} = \vec{r}_{in} \cdot \vec{n}$,
where $\vec{n}$ denotes the unit vector normal to the core-cladding
interface at the point of reflection and $\vec{r}_{in}=
(\vec{r}-\vec{r}_R)/|\vec{r}-\vec{r}_R|$ is the unit incident
propagation vector. Then, the reflection probability corresponding to
this angle $\alpha$ is determined. In case the ray is partially or
totally internally reflected, the total number of reflections is
increased and the unit propagation vector after reflection,
$\vec{r}_{\it out}$, is calculated by mirroring $\vec{r}_{in}$ with
respect to the normal vector: $\vec{r}_{\it out} = \vec{r}_{in} - 2
\vec{n} \cos{\alpha}$. The programme returns in a loop to the
calculation of the next reflection point. When the ray is absorbed on
its path or not reflected at the reflection point, the next ray is
generated. At any point of the ray's path axial, azimuthal and skew
angle are given by scalar products of the ray vector with the
coordinate axes in a projection on a plane perpendicular to the fibre
axis and parallel to the fibre axis, respectively. The transmitted
flux of a specific fibre, taking all losses caused by bending,
absorption and reflections into account, is calculated from the number
of lost rays compared to the number of rays reaching the fibre exit
end.

This method gives rise to an efficient simulation technique for fibres
with constant curvature. It is possible to extend the method for the
study of arbitrarily curved fibres by using small segments of constant
curvature. In the current version of the programme light rays are
tracked in the fibre core only and no tracking takes place in the
surrounding cladding, corresponding to infinite cladding thickness.
For simplicity only monochromatic light is assumed in the simulation
and highly wavelength-dependent effects are not included
explicitly. The simulation code takes about 1.5\,ms to track a skew
ray through a curved fibre.

\section{Light Yield---Trapping of Light}

In practical applications, the light yield is the most important
criterion for the design of a fibre detector.  Any deficiency in this
respect reduces detection efficiency, compromises timing resolution
and restricts total fibre length. Due to the small cross-section of
fibres the light yield is intrinsically low. Whether the fibres are
scintillating or wavelength-shifting one is only ever concerned with a
few 10's or 100's of photons propagating in the fibre and a single
photon counting capability is often necessary.  With increasing dopant
concentration the light yield increases, but above a certain limit
self-absorption of the emitted light reduces the attenuation length
noticeable. Accordingly, fibres with various concentrations of
fluorescent dyes have been systematically tested since 1992. For all
commercially available fibres an optimum has been found and no further
improvements from adjusting the dopant concentrations can be
expected. Thus, the trapped light as a fraction of the intensity of
the emitted light and the transmission function have the largest
impact on the light yield of an active fibre.

\subsection{Angular Phase Space of Trapped Light}

The geometrical path of any rays in cylindrical fibres, including skew
rays, was first analysed in a series of papers by
Potter~\cite{Potter1961} and Kapany~\cite{Kapany1961}. The treatment
of angular dependencies in this paper is based on their approach. The
angle $\gamma$ is defined as the angle of the projection of the light
ray in a plane perpendicular to the axis of the fibre with respect to
the normal at the point of reflection. One may describe $\gamma$ as a
measure of the ``skewedness'' of a particular ray, since meridional
rays have this angle equal to zero. The polar angle, $\theta^\prime$,
is defined as the angle of the light ray in a plane containing the
fibre axis and the point of reflection with respect to the normal at
the point of reflection. It can be shown that the angle of incidence
at the core-cladding interface of the fibre, $\alpha$, is given by
$\cos{\alpha}= \cos{\theta^\prime}\, \cos{\gamma}$. The values of the
two orthogonal angles $\theta^\prime$ and $\gamma$ will be preserved
independently for a particular photon at every reflection along its
path.

In general, for any ray to be internally totally reflected within the
fibre core, the inequality $\sin{\alpha} \geq \sin{\theta^\prime_{\it
crit}} = n_{\it clad}/n_{\it core}$ must be fulfilled, where the
critical angle, $\theta^\prime_{\it crit}$, is given by the index of
refraction of the fibre core, $n_{\it core}$, and that of the
cladding, $n_{\it clad}$. For the further discussion in this paper it
is convenient to use the axial angle, $\theta\ [= \pi/2 -
\theta^\prime]$, as given by the supplement of $\theta^\prime$, and
the skew angle, $\gamma$, to characterise any light ray in terms of
its orientation. A skew ray can be totally internally reflected at
larger angles $\theta$ than meridional rays and the relationship
between the minimum permitted skew angle, $\overline{\gamma}$, at a
given axial angle is determined by the critical angle condition:
$\cos{\overline{\gamma}}= \sin{\theta_{\it crit}} / \sin{\theta}$. In
the meridional approximation the above equations lead to the
well-known critical angle condition for the polar angle,
$\theta^\prime \ge \theta^\prime_{\it crit}$, which describes an
acceptance cone of semi-angle $\theta$ with respect to the fibre axis
(see for example~\cite{Potter1961} and references therein). Thus, in
this approximation all light within the forward cone, which
experiences multiple total internal reflections, will be considered as
trapped.

\begin{figure}[htbp]
  \begin{center}
    \subfigure[]{
	\epsfig{width= 0.47 \textwidth, file= 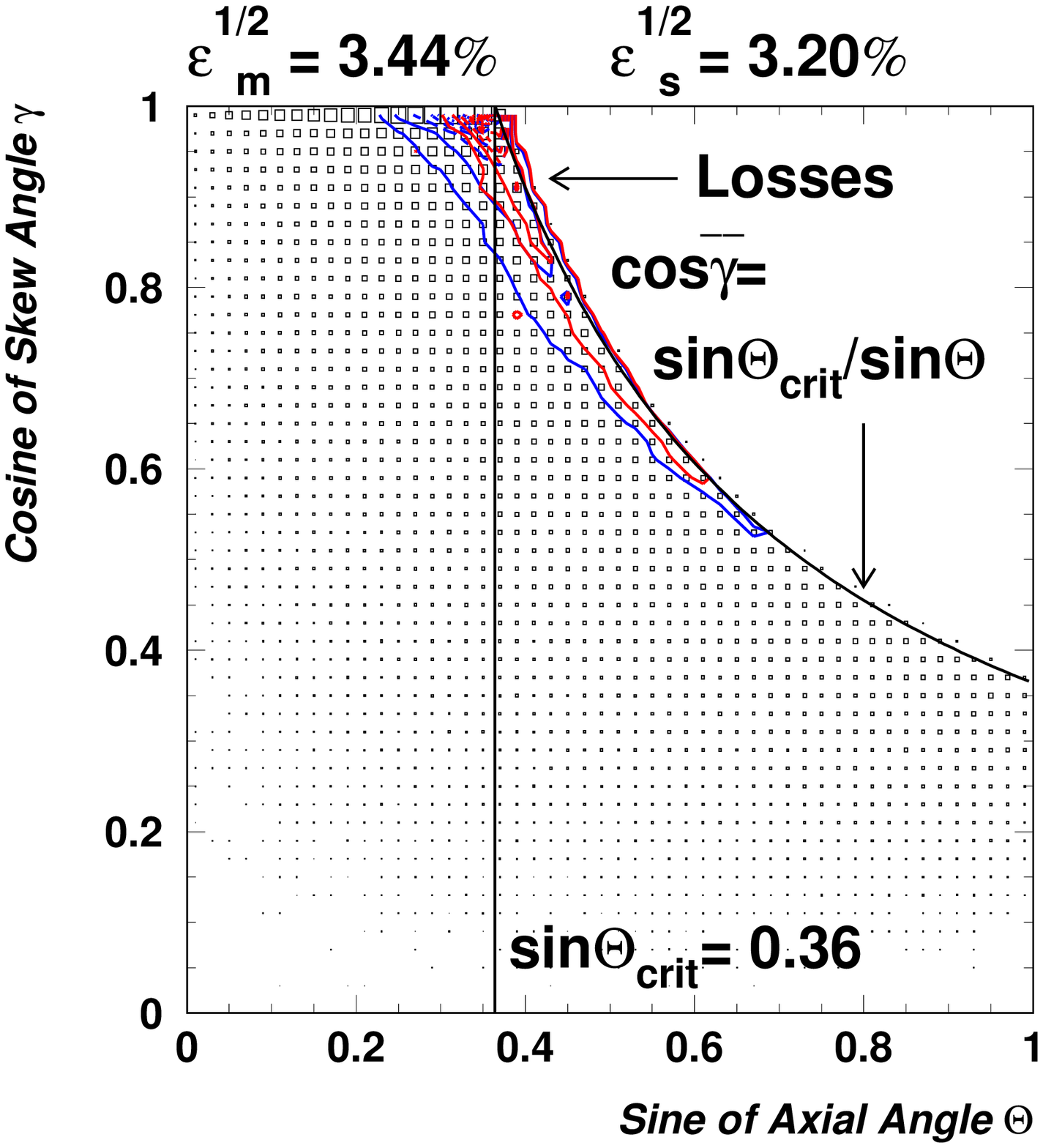} }
    \subfigure[]{
	\epsfig{width= 0.47 \textwidth, file= 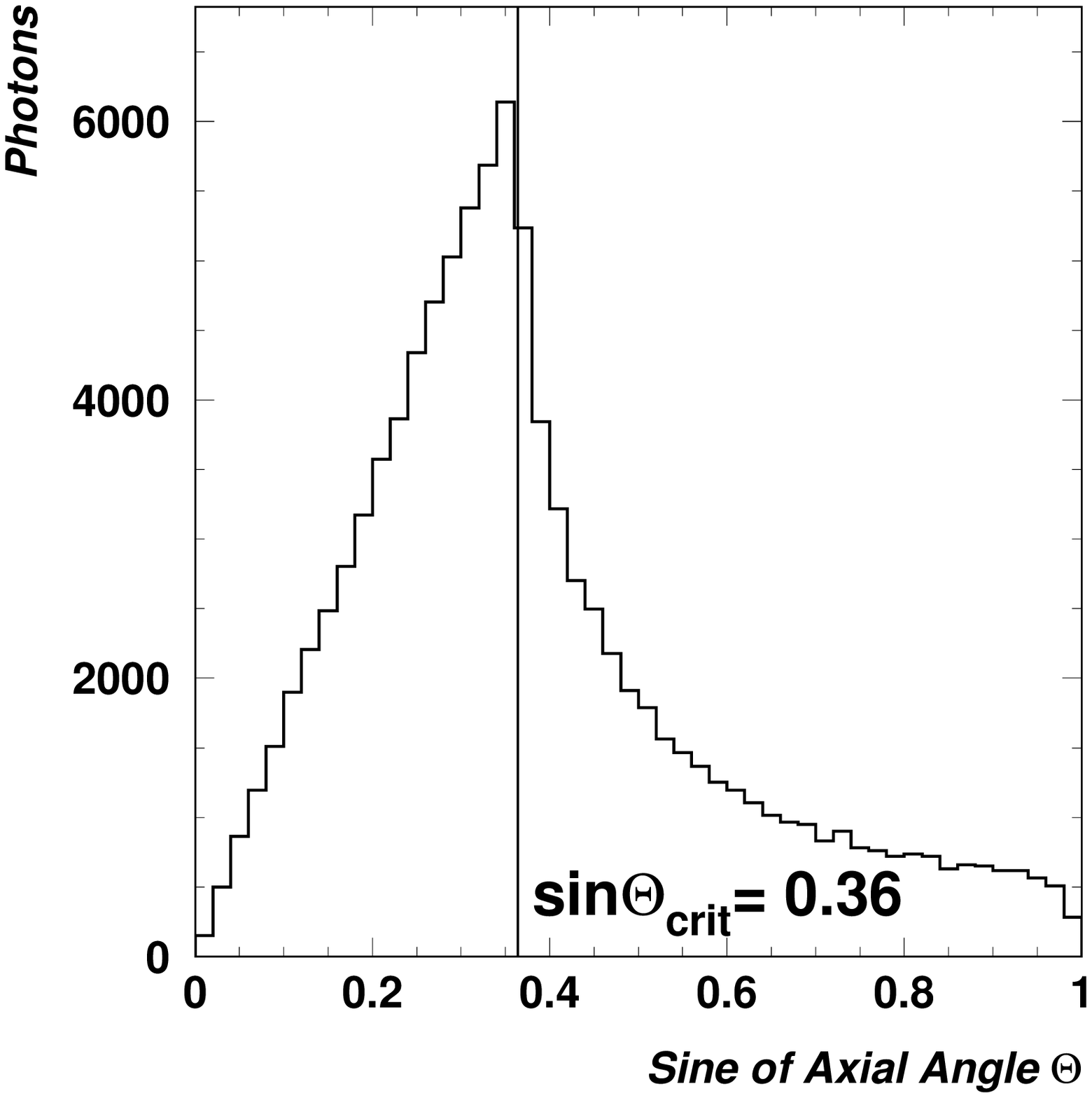} }\\[5mm]
    \caption{ (a) The angular phase space density for trapped photons 
	in a fibre. The given trapping efficiencies 
	are evaluated by integration over the two regions. The label
	``Losses'' points to contours for photons refracted 
	out of sharply curved fibres. (b) The projection of the 
	straight fibre phase space density onto the $\sin\theta$-axis.}
    \label{fig:phasespace}
  \end{center}
\end{figure}

Figure~\ref{fig:phasespace}(a) shows the total acceptance domain,
i.e.\ the angular phase space of possible propagation modes, and the
phase space density, i.e.\ the number of trapped photons per angular
element $d\cos\gamma\ d\sin\theta$, which is represented by
proportional boxes. Photons have been generated randomly on the
cross-section of the fibre with an isotropic angular distribution in
the forward direction. The density increases quadratically with
$\cos{\gamma}$ and linear with $\sin{\theta}$. To the left of the
dividing line at $\sin{\theta_{\it crit}}$ all skew angles are
accepted. To the right of the line a minimum skew angle is required by
the critical angle condition. The phase space contours in this figure
relate to sharply curved fibres. They show the distribution of photons
which are trapped in a straight fibre section but get refracted out of
sharply curved fibres with a radius of curvature to fibre radius
ratio, $R_{\it curv}/\rho$, of 33 and 83. The contours demonstrate
that only photons from a small region close to the phase space
boundary are getting lost. The smaller the radius of curvature, the
larger the affected phase space region. Figure~\ref{fig:phasespace}(b)
shows a projection of the phase space onto the $\sin\theta$-axis. A
peak around the value of $\sin{\theta_{\it crit}}$ is apparent.

\subsection{Trapping Efficiency}

The trapping efficiency for forward propagating photons,
$\epsilon^{\mathrm{fw}}$, may be defined as the fraction of totally internally
reflected photons. Figure~\ref{fig:phasespace}(a) gives values for the
two trapping efficiencies which have been determined by integrating
the phase space density over the two angular regions. The trapping
efficiency can also be determined analytically by two
integrals~\cite{Potter1963} for the flux transmitted by a fibre:
\begin{eqnarray}
  F & =  F_m + F_s
      =  & 4 \rho^2 \int_{\theta= 0}^{\theta_{\it crit}} 
	\int_{\gamma= 0}^{\pi/2} \int_{\phi= 0}^{\pi/2} 
	I(\theta,\phi)\, \cos^2{\gamma}\, d\gamma\, d\Omega\ + \nonumber \\
    & & 4 \rho^2 \int_{\theta= \theta_{\it crit}}^{\pi/2} 
	\int_{\gamma= \overline{\gamma}(\theta)}^{\pi/2} 
        \int_{\phi= 0}^{\pi/2}
	I(\theta,\phi)\, \cos^2{\gamma}\, d\gamma\, d\Omega\ ,
\label{eq:flux}
\end{eqnarray}
where $d\Omega$ is the element of solid angle,
$\overline{\gamma}(\theta)$ refers to the maximum axial angle allowed
by the critical angle condition, $\rho$ is the radius of the fibre and
$I(\theta,\phi)$ is the angular distribution of the emitted light in
the fibre core. The two integrals, $F_m$ and $F_s$, refer to the
meridional and skew case, respectively. The lower limit of the
integral $F_s$ is given by $\overline{\gamma}=
\arccos{(\sin{\theta_{\it crit}}/\sin{\theta})}$.

For an isotropic emission of fluorescence light the total flux through
the cross-section of the fibre core, $F_0$, equals $4 \pi^2 \rho^2
I_0$. Then, dividing the first term of equation~(\ref{eq:flux}) by the
total flux gives the trapping efficiency in the meridional
approximation,
\begin{equation}
  \epsilon^{\mathrm{fw}}_m = \frac{1}{2} (1 - 
	\cos{\theta_{\it crit}}) \approx 
	\frac{\theta^2_{\it crit}}{4}\ ,
  \label{eq:omega_m}
\end{equation}
where all photons are considered to be trapped if $\theta \le
\theta_{\it crit}$. Contributions of all skew rays to the trapping
efficiency are given by
\begin{equation}
  \epsilon^{\mathrm{fw}}_s = \frac{1}{2} (1 - \cos{\theta_{\it crit}})
  \cos{\theta_{\it crit}}\ .
  \label{eq:omega_s}
\end{equation}
The total trapping efficiency is then: 
\begin{equation}
  \epsilon^{\mathrm{fw}} = \frac{1}{2} (1 - \cos^2{\theta_{\it crit}}) 
	\approx \frac{\theta^2_{\it crit}}{2}\ ,
  \label{eq:omega_tot}
\end{equation}
which equals for small critical angles approximately twice the
trapping efficiency in the meridional approximation. This trapping
efficiency is crucially dependent on the circular symmetry of the
core-cladding interface. Any ellipticity or variation in the fibre
diameter will lead to the refraction of some skew rays.  Skew rays get
also attenuated more quickly, so that the trapping efficiency of skew
rays does not contribute to the light yield of a fibre in the same way
as the trapping efficiency of meridional rays does. In conclusion, for
long fibres the effective trapping efficiency is closer to
$\epsilon^{\mathrm{fw}}_m$ than to $\epsilon^{\mathrm{fw}}$.

The trapping efficiency for cladding light is determined by replacing
$n_{\it core}$ with $n_{\it clad}$ and $n_{\it clad}$ with the
refractive index of the surrounding medium, $n_{\it ext}$, in the
above formulae. Thus, the critical angle changes to $ \cos{\theta_{\it
crit}} = n_{\it ext}/n_{\it core}$. The cladding component of the
trapped light is the difference $\epsilon_{\it clad} = \epsilon -
\epsilon_{\it core} = \frac{1}{2} (n_{\it clad}^2 - n_{\it ext}^2)/
n_{\it core}^2$. This cladding light is usually a factor 3--4 more
intense than the core light. While this is highly desirable in terms
of light yield, it leads to a larger cross-talk between fibres of a
bundle. Furthermore, cladding light is heavily affected by the
external surface quality of the fibre: cracks in the cladding or
defects in the surface can cause significant light losses leading to
huge differences in the light yield of otherwise identical fibres.

Formula~\ref{eq:omega_m} gives a meridional ray trapping efficiency of
$\epsilon^{\mathrm{fw}}_m=$ 3.44\% for ``standard'' fibres with
$n_{\it core}=$ 1.6 and $n_{\it clad}=$ 1.49; the skew ray efficiency
in formula~\ref{eq:omega_s} evaluates to $\epsilon^{\mathrm{fw}}_s=$
3.20\% and the combined core efficiency in formula~\ref{eq:omega_tot}
to $\epsilon^{\mathrm{fw}}=$ 6.64\%. The efficiency for cladding light
neglecting any absorption equals 23.83\%. For square fibres the core
cladding efficiency is somewhat larger, the simulation calculates a
value of $\epsilon^{\mathrm{fw}}_{\mathrm{square}}=$ 8.13\%.

\begin{figure}[htbp]
  \begin{center}
    \epsfig{width= 0.5 \textwidth, file= 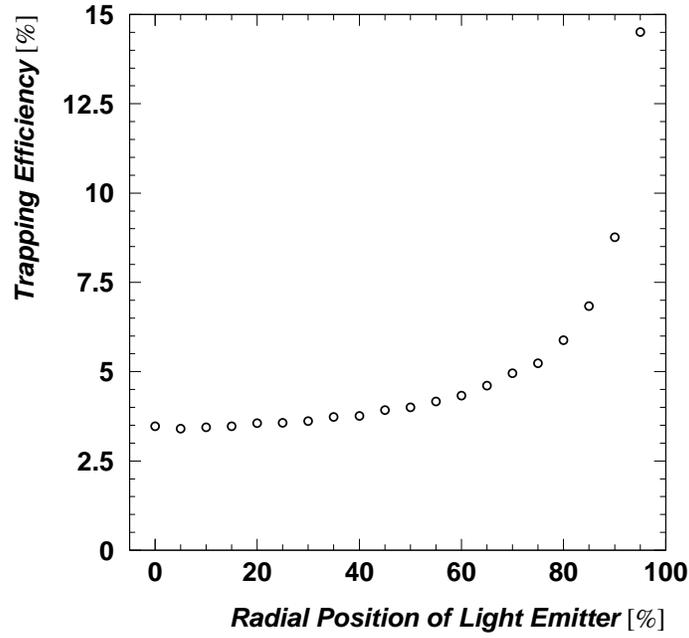}\\[5mm]
    \caption{ Trapping efficiency, $\epsilon^{\mathrm{fw}}$, for photons 
	propagating in the forward direction as a function of 
	radial position, $\hat{\rho}$, of the light emission point in 
	the fibre core.}
  \label{fig:trap-r}
  \end{center}
\end{figure}

It is obvious from the critical angle condition that a photon emitted
close to the cladding has a higher probability to be trapped than when
emitted close to the centre of the fibre. Scintillation photons are
distributed uniformly in solid angle. For a given axial angle the
range of possible azimuthal angles for the photon to get trapped
increases with the radial position, $\hat{\rho}$, of the light
emission point in the fibre core. Figure~\ref{fig:trap-r} shows the
core trapping efficiency, $\epsilon^{\mathrm{fw}}$, for photons
propagating in the forward direction as a function of radial position,
$\hat{\rho}$, of the emission point in the fibre core. It can be
deduced from figure~\ref{fig:trap-r} that the meridional approximation
is a good estimate for $\epsilon$ if the photons originate at radial
positions $\hat{\rho} < 0.8$. The trapping of skew rays only becomes
significant for photons originating at radial positions $\hat{\rho}
\ge 0.9$. This fact has been discussed e.g.\ in~\cite{Johnson1994}.

The maximum possible concentration factor of an optical system without
any light loss is $1/\sin^2\theta$, where $\theta$ describes the
divergence of light in the system. This angle can be
approximated by the maximum axial angle of trapped light in active
elements. In fibres, the comparatively small angular phase space
permits the use of optical concentrators with high concentration
factors, and the concentration is possible by means of total internal
reflections within a light guide. Optical concentrators, mostly of the
Winston type, have been built to couple scintillation light
efficiently to photo-detectors~\cite{Kuhlen1991}.

\subsection{Transmission of Straight Fibres}

A question of practical importance for the estimation of the light
output of a particular fibre application is its transmission function,
which quantifies the transmission probability of trapped photons. The
function is dependent on the total photon path length per axial fibre
length, $P$, the number of internal reflections per axial fibre
length, $\eta$, and the optical path length between successive
internal reflections, $l_R$. It should be noted that these three
variables are not independent as $P= \eta \times l_R$.

Light attenuation in active fibres has many sources, among them
absorption in the base material, at optical non-uniformities or at
impurity centres, as well as reflection losses caused by a rough
surface or variations in the refractive indices. Skew light is
attenuated by stronger absorption and reflection losses than
meridional light because of the longer path length and the higher
number of reflections it suffers from.  Accordingly, the light
attenuation at short distances differs from the attenuation at large
distances.  Furthermore, the absorption and emission processes in
fibres are spread out over a wide band of wavelengths and the
attenuation is known to be wavelength dependent. The attenuation of
active fibres at wavelengths close to its emission band
($400-600$\,nm) is much higher than in wavelength regions of interest
for communication applications where mainly infrared light is
transmitted ($0.8-0.9\,\mu$m and $1.2-1.5\,\mu$m).

The two main sources of attenuation in the base material are
self-absorption of scintillation light and Rayleigh scattering. The
cumulative effect of these attenuation processes can be conveniently
parameterised by an effective attenuation length, $\Lambda_{\it eff}$,
over which the signal amplitude is attenuated to 1$/e$ of its original
value. This parameter is often used in particle physics applications
to characterise the attenuation, but is of limited significance when
analysing different causes and effects of attenuation. Instead, the
transmission function of a straight fibre should be written as a
function of the axial angle
\begin{equation}
  T(\theta) = {\mathrm e}^{- P(\theta)\, L_F/\Lambda_{\it bulk}}
  \times q^{\eta(\theta) L_F}\ ,
\end{equation}
where the bulk attenuation length $\Lambda_{\it bulk}^{-1} =
\Lambda_{\it abs}^{-1} + \Lambda_{\it scat}^{-1}$ describes light
losses due to bulk absorption (bulk absorption length $\Lambda_{\it
abs}$) and scattering (scattering length $\Lambda_{\it scat}$), and
the second factor describes light losses due to imperfect reflections
(reflection coefficient $q$ to parameterise the internal
reflectivity). The self-absorption is mainly caused by an overlap of
the absorption and emission bands of the fluorescent dyes. The
scattering length quantifies Rayleigh scattering on small density
fluctuations in the core. The cross-section for the scattering
processes increases with decreasing wavelength and becomes noticeable
in the region of the emission peak of the base scintillator (200 --
300\,nm). A comparison of available data indicates that a reasonable
value of the bulk attenuation length in polystyrene is $\Lambda_{\it
bulk} \simeq 3 - 5$\,m for doped fibres and $\simeq 8$\,m for clear
fibres. Fibres are drawn from a boule and great care is taken during
production to ensure that the core-cladding interface has the highest
possible uniformity and quality. Most published data suggest a
deviation of the reflection coefficient from unity between $5 \times
10^{-5}$ and $6.5 \times 10^{-5}$ \cite{Ambrosio1991}. A reasonable
value of $q= 0.9999$ is used in the simulation to account for all
losses proportional to the number of reflections.

The light from a fibre may be read-out from either one or both ends.
A reflector at the open end of the fibre allows the collection of
photons propagating in direction opposite to the photon sensor which
would normally escape from the fibre. Mirroring may be applied by
sputtering an aluminium coating onto the fibre end or by bringing the
fibre end into direct contact with a highly specular reflector such as
an aluminised mylar foil. Simple foils provide a reflectivity $R
\approx 70\%$ and can lead to an increase in light yield of
20\%. Covering the fibre end with a white diffuse reflector can help,
as well. When describing the light yield of these fibres a second term
has to be added to the transmission function to account for the
reflected light. The proportion of the direct, $T_d$, to the
reflected, $T_r$, light intensity depends on the distance to the
emission point, $L_0$, and the reflection coefficient, $R$, of the
reflector. The transmission function becomes:
\begin{equation}
  T = T_{d} + R\, T_{r} = {\mathrm e}^{- P\, L_0/\Lambda_{\it bulk}}
  \times q^{\eta L_0}\ + R \left( {\mathrm e}^{- P\, (2L_F - L_0
  )/\Lambda_{\it bulk}} \times q^{\eta (2L_F - L_0)} \right)\\ .
\end{equation}
Comparison of the signal arrival times from each end can be used to
determine the longitudinal position of the light emission. For
simplicity, in this paper the analysis and discussion is restricted to
the direct light only.

\begin{figure}[htbp]
  \begin{center}
    \epsfig{width= 0.5 \textwidth, file= 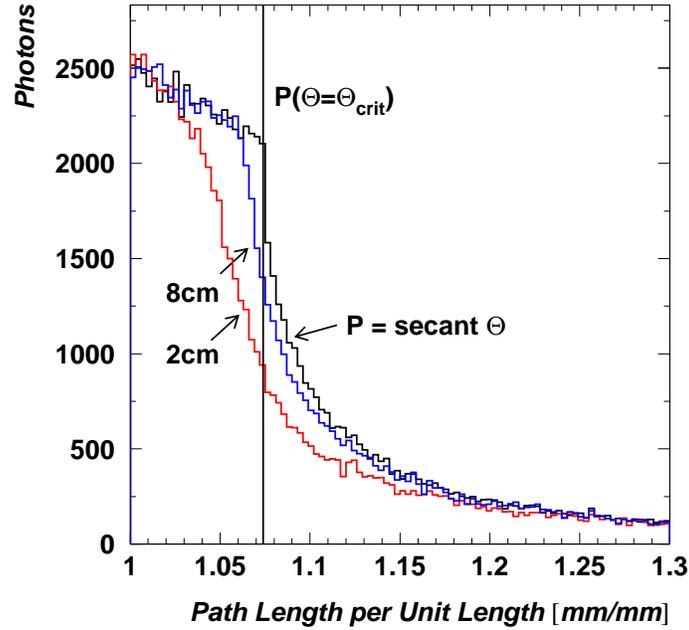}\\[5mm]
    \caption{ The distribution of the optical path length, $P(\theta)$, 
	of trapped photons in fibres of radius $\rho=$ 0.6\,mm 
	normalised to the axial length of the fibre. The figure shows 
	$P(\theta)$ for a straight fibre and for two different radii 
	of curvature, $R_{\it curv}=$ 2 and 8\,cm. The vertical line at 
	$P(\theta_{\it crit})=$ 1.074 indicates the upper limit of 
	$P$ in the meridional approximation.}
    \label{fig:pathlength}
  \end{center}
\end{figure}

Figure~\ref{fig:pathlength} shows the distribution of the normalised
path length, $P(\theta)$, for trapped photons reaching the exit end of
straight fibres of 0.6\,mm radius. The figure also gives results for
curved fibres of two different radii of curvature. The distribution of
path lengths which are shorter than the path length for meridional
photons propagating at the critical angle is almost flat. It can be
shown that the normalised path length along a straight fibre is given
by the secant of the axial angle and is independent of other fibre
dimensions: $P(\theta)= \sec\theta$. It is clearly seen that, when a
fibre is curved, the normalised path length of the trapped photons is
less than the secant of the axial angle and photons on near meridional
paths are refracted out of the fibre most. The average normalised path
length for those photons which remain trapped is smaller than the
average for the straight fibre. The over-all fibre length for the
curved fibres in these calculations is 0.5\,m and the fibres are
curved over a circular arc for their entire length.

\begin{figure}[htbp]
  \begin{center}
    \epsfig{width= 0.5 \textwidth, file= 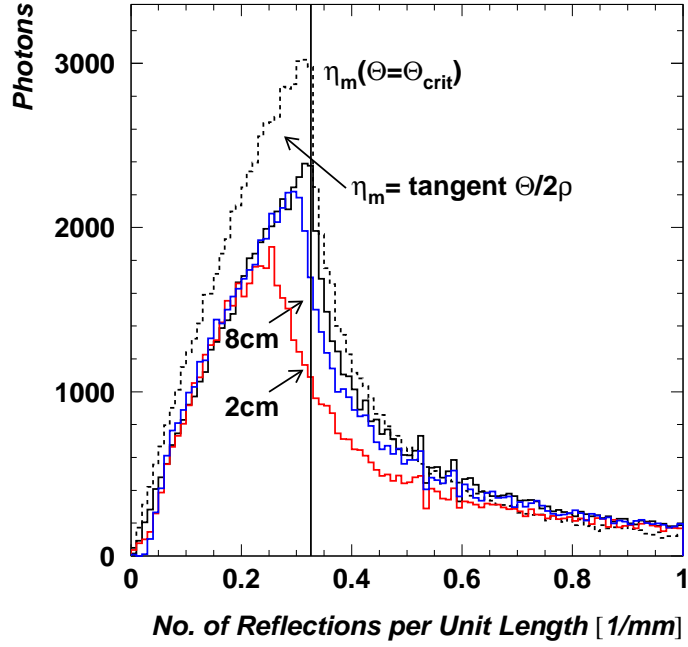}\\[5mm]
    \caption{ The distribution of the number of reflections, 
	$\eta(\theta)$, for trapped photons in fibres of radius 
	$\rho=$ 0.6\,mm normalised to the axial length of the fibre. 
	The figure shows $\eta(\theta)$ for a straight fibre and 
	for two different radii of curvature, $R_{\it curv}=$ 2 
	and 8\,cm. The vertical line at $\eta_m (\theta_{\it crit})=$ 
	0.326 indicates its upper limit in the meridional approximation. 
	The dashed line shows the distribution of 
	$\eta_m(\theta)= \tan{\theta}/2\rho$.}
    \label{fig:reflections}
  \end{center}
\end{figure}

The distribution of the normalised number of reflections,
$\eta(\theta)$, for photons reaching the exit end of straight and
curved fibres is shown in figure~\ref{fig:reflections}. Again, the
figure gives results for curved fibres of two different radii of
curvature. The number of reflections a photon experiences scales with
the reciprocal of the fibre radius. In the meridional approximation
the normalised number of reflections is related by simple trigonometry
to the axial angle and the fibre radius: $\eta_m(\theta) =
\tan{\theta}/2\rho$. The distribution of $\eta_m$, based on the
distribution of axial angles for the trapped photons, is represented
by the dashed line. The upper limit, $\eta(\theta_{\it crit})$, is
indicated in the plot by a vertical line. The number of reflections
made by a skew ray, $\eta_s(\theta)$, can be calculated for a given
skew angle: $\eta_s(\theta)= \eta_m(\theta) / \cos{\gamma}$. It is
clear that this number increases significantly if the skew angle
increases. From the distributions it can be seen that in curved fibres
the trapped photons experience fewer reflections on average.

Internal reflections being less than total give rise to so-called
leaky or non-guided modes, where part of the electromagnetic energy is
radiated away. Rays in these modes populate a region defined by axial
angles above the critical angle and skew angles slightly larger than
the ones for totally internally reflected photons. In the simulation
these modes are taken into account by using the Fresnel equation for
the reflection coefficient, $\langle R \rangle$, averaged over the
parallel and orthogonal plane of polarisation
\begin{equation}
  \langle R \rangle = \frac{1}{2} \left( R_{||} + R_\perp \right) = 
     \frac{1}{2} \left( \frac{\tan^2(\alpha - \beta)}
     {\tan^2(\alpha + \beta)} + \frac{\sin^2(\alpha - \beta)}
     {\sin^2(\alpha + \beta)} \right)\ ,
\end{equation}
where $\alpha$ is the angle of incidence and $\beta$ is the refraction
angle. However, it is obvious that non-guided modes are lost quickly
in a small fibre. This is best seen in the fraction of non-guided to
guided modes, $f$, which decreases from $f = 11\%$ at the first
reflection of the ray over $f = 2.5\%$ at the second reflection to $f
< 1\%$ at further reflections. Since the average reflection length of
non-guided modes is $l_R \approx 1.5$\,mm those modes do not
contribute to the flux transmitted by fibres longer than a few
centimetres.
 
\begin{figure}[htbp]
  \begin{center}
      \epsfig{width= 0.5 \textwidth, file= 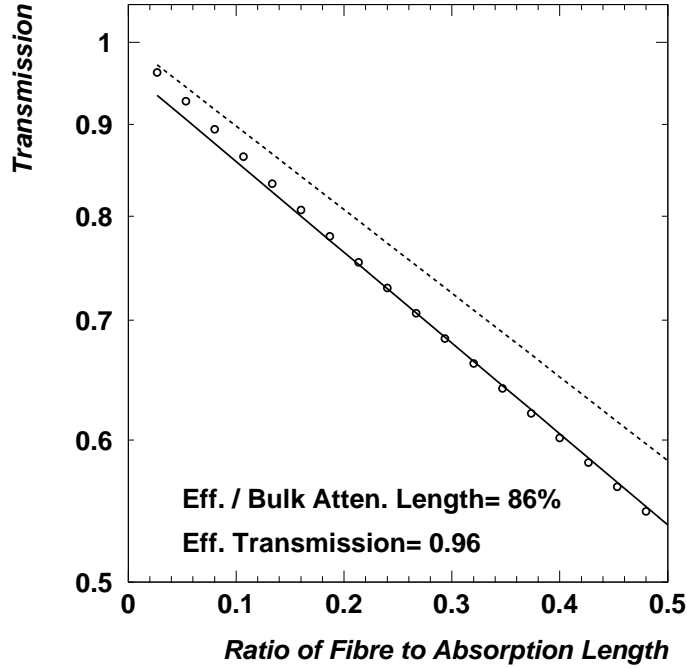}\\[5mm]
      \caption{ Transmission function for a straight fibre as a function 
	of the ratio of fibre to absorption length. The open circles
	are calculated from the optical path length distribution of the 
	simulation. A simple exponential fit results in an effective 
	attenuation length $\Lambda_{\it eff}= 86\%\ \Lambda_{\it bulk}$. 
	The dashed line shows the transmission function in the 
	meridional approximation with $\Lambda_m= 93\%\ \Lambda_{\it bulk}$.}
      \label{fig:absorption}
  \end{center}
\end{figure}

In the meridional approximation and substituting $\exp(-\ln{q})$ by
$\exp(1-q)$ the attenuation length can be written as
\begin{equation}
  \Lambda_m = \cos{\theta_{\it crit}}\, \left[ 1/\Lambda_{\it bulk} +
  (1-q)\sin{\theta_{\it crit}}/2\rho \right]^{-1}\ .
\end{equation}
Only for small diameter fibres ($2\rho \le 0.1\,$mm) the attenuation
due to imperfect reflections is of the same order as the absorption
lengths. In calorimeters the reflection losses are not relevant for
the transmission function, because of the large radii of the fibres
used. There, the attenuation length contracts to $\Lambda_m=
\Lambda_{\it bulk} \cos{\theta_{\it crit}}$. The transmission function
including all skew ray effects can be found by integrating over the
normalised path length distribution
\begin{equation}
  T = \frac{1}{N} \int_{P=0}^{\infty} e^{-P\, L_F/
	\Lambda_{\it bulk}}\, \frac{dN}{dP}\, dP\ ,
\end{equation}
where $dN$ represents the number of photons per path length interval
$dP$, weighted by the exponential bulk attenuation
length. Figure~\ref{fig:absorption} shows the simulated transmission
function versus the ratio of fibre to absorption length,
$L_F/\Lambda_m$. A simple exponential fit, $T \propto
\exp\left[-L_F/\Lambda_{\it eff}\right]$, applied to the simulated
light transmissions for a large number of fibre lengths results in an
effective attenuation length of $\Lambda_{\it eff}= 86\%\ \Lambda_{\it
bulk}$. For $L_F/\Lambda_m \ge 0.2$ this number is sufficiently
accurate to parameterise the transmission function, at smaller values
for $L_F/\Lambda_m$ the light is attenuated faster. The difference to
the meridional attenuation length, $\Lambda_m= 93\%\ \Lambda_{\it
bulk}$, is attributed to the skew rays in the tail of the path length
distribution.

Measurements of the light attenuation in fibres proved the simple
model of a single effective attenuation length to be wrong. A
dependence of the attenuation length with distance is usually
observed~\cite{Davis1989}. One cause of this effect is the fact that
the short wavelength components of the scintillation light is
dominantly absorbed. The use of a double spectrometer allows the
precise determination of the spectral attenuation length of
fibres~\cite{Drexlin1995}. The characteristic shape of plastic
spectral attenuation lengths shows a minimum at around 440\,nm and an
increase towards longer wavelengths. This leads to a shift of the
average wavelength in the emission spectrum towards longer wavelengths
and to an increase in the effective attenuation length. The wavelength
dependent quantum efficiency of any photon sensor enhances this
effect, so that the integral attenuation length is only a rough
quality criterion for the light yield of a fibre.

\subsection{Transmission of Sharply Curved Fibres}

One of the most relevant practical issues in implementing optical
fibres into compact particle detector systems are macro-bending
losses. In general, some design parameters of fibre applications,
especially if the over-all size of the detector system is important,
depend crucially on the minimum permissible radius of curvature. The
problem of bending is eminent for tile-fibre calorimeters where
wavelength-shifting fibres are embedded in plastic scintillator tiles
and are bent at radii of curvature of a few centimetres.  Flexibility
of the fibres is also essential for space physics experiments or for
detectors carried by balloons or aircrafts. The routing of fibres to
photon sensors requires radii of curvature as small as possible to
minimise the weight and the costs associated with the transport of the
detector, see e.g.~\cite{Adler2001}.

Photons are lost from a fibre core both by refraction and
tunnelling. In the simulation only refracting photons were considered.
The angle of incidence of a light ray at the tensile (outer) side of
the fibre is always smaller than at the compressed side and photons
propagate either by reflections on both sides or in the extreme
meridional case by reflections on the tensile side only. If the fibre
is curved over an arc of constant radius of curvature, photons can be
refracted, and will then no longer stay trapped, at the very first
reflection point on tensile side. Therefore, the trapping efficiency
for photons entering a curved section of fibre towards the tensile
side is reduced most. 

\begin{figure}[htbp]
  \begin{center}
    \epsfig{width= 0.5 \textwidth, file= 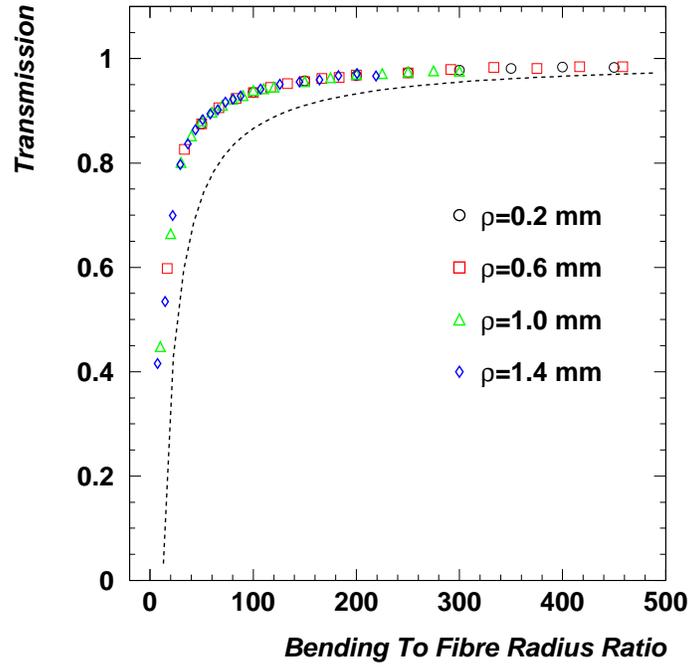}\\[5mm]
    \caption{ The transmission function for fibres curved over a circular
	arc of 90\,$^\circ$ is plotted as a function of the radius of 
    	the curvature to fibre radius ratio for different fibre radii, 
	$\rho=$ 0.2, 0.6, 1.0 and 1.4\,mm. The dashed line is a simple
	estimate from the meridional approximation.}
    \label{fig:bradius}
  \end{center}
\end{figure}

Figure~\ref{fig:bradius} displays the explicit dependence of the
transmission function for fibres curved over circular arcs of
90$^\circ$ on the radius of curvature to fibre radius ratio for
different fibre radii, $\rho=$ 0.2, 0.6, 1.0 and 1.2\,mm. No further
light attenuation is assumed. Evidently, the number of photons which
are refracted out of a sharply curved fibre increases very rapidly
with decreasing radius of curvature. The losses are dependent only on
the curvature to fibre radius ratio, since no inherent length scale is
involved, justifying the introduction of this scaling variable. The
light loss due to bending of the fibre is about 10\% for a radius of
curvature of 65 times the fibre radius.

The transmission function in the meridional approximation in the
bending plane can be estimated by assuming that all photons with axial
angles above a limiting angle, $\theta_0 (\theta, R_{\it curv}/\rho$),
are refracted out of the fibre:
\begin{equation}
  T= 1 - \frac{1}{1 + R_{\it curv}/\rho}\ 
  \frac{\cos\theta_{\it crit}}{1 - \cos\theta_{\it crit}}\ .
\end{equation}
This transmission function is shown in figure~\ref{fig:bradius} as a
dashed line to be compared with the simulation results including all
skew rays.  It overestimates the light losses due to the larger axial
angles allowed for skew rays.  The meridional approximation in the
bending plane is, however, a good approximation for the transmission
function because the light losses are dominantly produced by
meridional rays~\cite{Winkler1979,Gloge1972}.

A comparable theoretical calculation using a two-dimensional slab
model and a generalised Fresnel transmission coefficient has been
performed by Badar and co-workers~\cite{Badar1991A}. Their plot of the
power contained in the fibre core as a function of the radius of
curvature (figure~5 in~\cite{Badar1991A}) shows results on the
transmission function in the meridional approximation which are
similar to the simulation output. In~\cite{Winkler1979} a ray optics
calculation for curved multimode fibres involving skew rays is
presented, unfortunately a discussion on the transmission function is
missing. Instead, a plot of the power remaining in a curved fibre
versus distance is shown which gives complementary information.

\begin{figure}[htbp]
  \begin{center}
      \epsfig{width= 0.5 \textwidth, file= 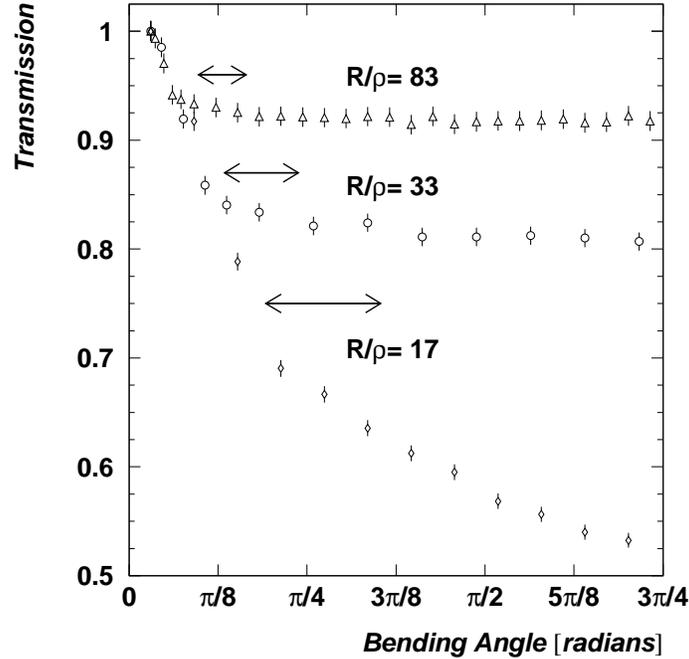}\\[5mm]
      \caption{ Transmission function for a curved fibre 
	of radius $\rho=$ 0.6\,mm with three different radii of 
        curvature, $R_{\it curv}=$ 1, 2 and 5\,cm, corresponding to 
	the ratios $R_{\it curv}/\rho=$ 17, 33 and 83, respectively. 
	The ordinate shows the fraction of photons reaching the fibre exit 
	end as a function of the bending angle, $\Phi$, and the arrows 
	indicate the transition region in the meridional approximation.}
      \label{fig:bending}
  \end{center}
\end{figure}

\begin{figure}[htbp]
  \begin{center}
    \epsfig{width= 0.4 \textwidth, file= 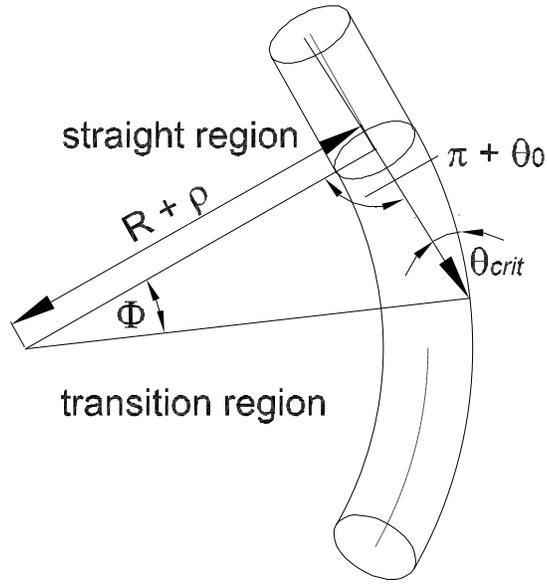}\\[5mm]
    \caption{ Section of a curved fibre with radius $\rho$ and radius 
	of curvature $R_{\it curv}$. The passage of a meridional ray 
	in the bending plane with maximum axial angle is shown.}
    \label{fig:bentfibre}
  \end{center}
\end{figure}

For photons entering a curved section of fibre the first point of
reflection on the tensile side defines the transition angle,
$\Phi_{\it trans}$, measured from the plane of
entry. Figure~\ref{fig:bentfibre} shows a section of a curved fibre
and the passage of a meridional ray in the bending plane with maximum
axial angle. The angular range of transition angles associated with
each ray is called the transition region of the fibre. The simulation
results on the transmission as a function of bending angle, $\Phi$,
for a ``standard'' fibre are presented in
figure~\ref{fig:bending}. Once a sharply curved fibre with a ratio
$R_{\it curv}/\rho > 83$ is bent through angles $\Phi \simeq
\pi/8$\,rad, light losses do not increase any further. For ratios
$R_{\it curv}/\rho$ smaller than 10 the model is no longer valid to
describe the transmission function.

For photons emitted towards the tensile side the transition angle is
related to the axial angle. Since the angular phase space density of
trapped photons is highest close to the critical angle the difference
$\theta_{\it crit} - \theta_0$ allows to estimate the transition
angle. Photons emitted from the fibre axis towards the compressed side
are not lost at reflections on this side. They experience at least one
reflection on the tensile side if the bending angle exceeds the limit
$\Phi_{\it limit} = \arccos\left[R_{\it curv}/(R_{\it curv} + 2\,
\rho)\right] \approx \arccos\left[1 - 2\, \rho / R_{\it
curv}\right]$. Therefore, a transition in the transmission function
should occur at bending angles between $\Phi_{\it limit}/2$, where
photons emitted towards the tensile side must have experienced a
reflection, and $\Phi_{\it limit}$, where this is true for all
photons. For a fibre radius $\rho=$ 0.6\,mm and radii of curvature
$R_{\it curv}=$ 1, 2 and 5\,cm the above formula leads to transition
regions $\Phi \sim 0.44 - 1.06$\,rad which are indicated in
figure~\ref{fig:bending} by arrows.

Transition and bending losses have been thoroughly investigated using
waveguide analysis techniques from which a loss formula in terms of
the Poynting vector has been
derived~\cite{Marcuse1976,Gambling1979}. Those studies are difficult
to extend to multimode fibres since a large number of guided modes has
to be considered. Despite the extensive coverage of theory and
experiment in this field, only fragmentary studies on the trapping
efficiencies and refraction of skew rays in curved multimode fibres
could be found~\cite{Winkler1979,Badar1991A,Badar1989,Badar1991B}.
When applying ray optics to curved multimode fibres the use of a
two-dimensional model is very
common~\cite{Badar1991A,Badar1989,Badar1991B}. In contrast, the
simulation method presented in this paper follows a three-dimensional
approach.

Experimental results on losses in curved multimode fibres along with
corresponding predictions are best known for silica fibres with core
radii $\rho \approx 50\,\mu$m. Results on multimode plastic fibres are
rare. The manufacturer Kuraray has investigated a time dependent drop
of the transmission of curved plastic fibres on time scales of several
days~\cite{Hara1998}. A dependence of the bending losses on the $S$
parameter was found. This parameter is a characteristic index for the
degree of orientation of polystyrene chains along the fibre axis,
where fibres are more flexible for larger $S$ values. Consequently,
softer and more flexible fibres have been developed by Kuraray
(so-called $S$ type fibres). Bending losses of more than 50\% in 1\,mm
diameter Bicron BCF-91A fibres have been observed for one turn of
360$^\circ$ with a radius of curvature of 5\,cm~\cite{Gomes1998}. With
the same type of fibres no effect was seen for a radius of
10\,cm. Calculations on the basis of ray optics for a plastic fibre
with $\rho = 0.49\,$mm can be found in~\cite{Badar1991B}. The
simulation result on the transmission function in the meridional
approximation $T= 0.35$ at $\rho/R_{\it curv}= 20$ is in good
agreement with this two-dimensional calculation. The higher
transmission of $T= 0.65$ for this curvature as predicted by the
simulation is explained by the fact that skew rays are less sensitive
to bending effects. It should be noted that the difference between
finite and infinite cladding and oscillatory losses in the transition
region have not been investigated in the simulation.

\subsection{Transmission of Irradiated Fibres}

It is frequently desirable to have fibre detectors very close to beams
and targets. This task results in rather demanding specifications. The
fibres have to reasonably immune to radiation or any other effect of
aging within the expected period of operation.

Radiation resistance remains to be an important issue for detectors at
high luminosity accelerators. It is influenced by many parameters:
dose rate, recovery times, temperature, chemical composition, level of
dissolved oxygen and others. Since the effective attenuation length is
essentially the only property which is affected by the irradiation,
the transmission curves before and after exposure to electromagnetic
radiation can be used to study the influence of the deposited energy
dose on the light yield. Many data points for various irradiation
conditions exist. From these experiences one can draw the following,
simplified conclusions for most modern fibres: after irradiation the
attenuation length is reduced significantly compared to the original
value, while the scintillation mechanisms themselves are not strongly
affected. Parts of the attenuation length recover on multiple time
scales. The manufacturer Kuraray gives a simple formula for estimating
the ratio of attenuation lengths after to before the irradiation as a
function of the absorbed dose, $D$ (for $D > 0.1$), in units of krad
(1\,krad $=$ 10\,Gy)~\cite{Hara1998}:
\begin{equation}
  \Lambda/\Lambda_0 = 0.80 \pm 0.01 - (0.144 \pm 0.007) \times \log D\ .
\end{equation}

In general, strong dose rate effects are observed and a more accurate
description has to include the dynamics of absorption centres. Three
types of absorption centres can be formed during irradiation: (1)
stable absorption centres leading to a permanent damage, (2) radicals
(e.g.\ benzyl radicals of polystyrene) which decay practically at once
if oxygen is dissolved in the fibre, and (3) short-lived absorption
centres which decay within hours via bi-molecular
reactions~\cite{Wick2001}. The short-lived absorption centres in
polystyrene mainly absorb red and green light and it was concluded
that these centres are caused by the dopants and not the base
material.

For the quantitative analysis of the radiation induced changes the
difference in the absorption coefficients of the irradiated and
non-irradiated fibre, $\Delta k(\lambda) = k_{\it after}(\lambda) -
k_{\it before}(\lambda)$, is usually quoted, where $\lambda$ is the
wavelength and $k = \Lambda^{-1}$ is the inverse light attenuation
length. For many fibres the permanent induced absorption, $\Delta
k_{\it perm}$, that is the damage after the end of the recovery
process, rises linearly with the absorbed dose for moderate doses
between 0.1 and 6\,kGy~\cite{Wick2001}. For larger absorbed doses
$\Delta k_{\it perm}(D)$ becomes a non-linear function. The annealable
part of the absorption, $\Delta k - \Delta k_{\it perm}$, shows a
broad maximum in the blue spectral region of many irradiated fibres.

\section{Timing Properties}

The timing properties of active fibres are usually defined in terms of
a coincidence timing resolution between identical counters. For small
counters a timing resolution in the range of $\sigma_{tot} \approx 50
- 100$\,ps can be achieved. Several effects which are dependent on the
signal amplitude, $A$, contribute to the timing resolution, namely the
scintillation process with its decay time $\sigma_{sci}$, the
photomultiplier tube with its transit time spread $\sigma_{TTS}$ and
the fibre as a light guide with its pulse dispersion $\sigma_{disp}$.
The discriminator, which could be either of type leading edge or
constant fraction, and any noise in the electronic circuits contribute
to the timing distribution. They may shift the response time with
signal amplitude (``time walk''). Time walk effects can get corrected
by various means, either in hardware or
software~\cite{Spieler1982}. Usually, the remaining effect is
parameterised by the time spread $\sigma_{\it elec}$. A common way of
summarising the different contributions to the timing resolution is
the following:
\begin{equation}
  \sigma= \sqrt{2} \sqrt{\frac{\sigma_{\it sci}^2}{A} +
  \frac{\sigma_{\it TTS}^2}{A} + \frac{(L_F \sigma_{\it disp})^2}{A} +
  \sigma_{\it elec}^2}\ ,
\end{equation}
where the factor $\sqrt2$ assumes two identical counters for start and
stop time, whose timing resolutions add up quadratically: $\sigma^2 =
\sigma^2_{\it start} + \sigma^2_{\it stop}$.

\subsection{Statistical Fluctuations}

Obviously, the signal amplitude depends on the number of
photoelectrons, $N$, appearing during some integration time, $T$.
This number fluctuates from one pulse to another. The mean number of
photoelectrons per pulse, $\bar{N}$, and its variance, $\sigma_N$, are
characteristic of the photoelectron statistics. As the number of
photoelectrons increases, a larger number of time intervals is
sampled. For a precise quantitative description of the distribution of
$N$ one has to use quantum theory, where the number of photoelectrons
becomes an operator. However, in a semiclassical model the probability
for observing $N$ photoelectrons over a time interval $T$ is given by
a Poissonian distribution $P(\bar{N}, N)$.  For larger numbers of
photoelectrons the distribution is assumed to be Gaussian, so that the
signal amplitude dependent contributions should vary with
$1/\sqrt{\bar{N}}$.

In active fibres the addition of fluorescent dopants can sharply
reduce the decay times of some scintillating base materials by
F\"orster transitions, which couple base and fluorescent dye in
extremely short times. The base material polyvinyltoluene can be
quenched with benzophenone, for example, to reach a 220\,ps pulse
width. The faster timing comes at the expense of light yield,
however. For most plastic fibres the short time behaviour can be
described phenomenologically by a single decay constant,
$\sigma_{sci}$, usually of the order of a few nanoseconds. The decay
times in wavelength-shifting fibres are substantially longer than the
scintillator decay constants. Typical wavelength-shifting molecules
have decay times of $7-12$\,ns, dominating the time characteristics of
the entire detector. For critical timing situations
wavelength-shifting fibres are usually avoided.

The contribution of the photomultiplier tube to the time width of the
observed output pulse is determined exclusively by the electron
transit time spread, $\sigma_{TTS}$, where the transit time is the
time difference between photoemission at the cathode and the arrival
of the subsequent electric signal at the anode. Modern compact
photomultiplier tubes like the Hamamatsu R-5900 reach transit time
spreads of $\sigma_{TTS} \approx 100$\,ps with anode pulse rise times
of $600$\,ps for single photoelectrons. The timing performance of
photomultiplier tubes is an active area of development and it is
assumed that even faster tubes will be available. In many cases, the
transit time spread of conventional 1$\frac{1}{8}$ inch
photomultiplier tubes is well described by a $1/\sqrt{\bar{N}}$
dependence. Sometimes, other dependences of $\sigma_{TTS}$ on
$\bar{N}$ are seen. Ref.~\cite{Kume1986} gives an example of a
$\sigma_{TTS} \propto 1/\bar{N}^{0.4}$ dependence for a specific
photomultiplier tube.

\subsection{Light Pulse Dispersion}

A pulse of light, consisting of several photons propagating along a
light guide, broadens in time. The chromatic dispersion is due to the
spectral width, $\Delta\lambda$, of the emission band. It is a
combination of material dispersion and waveguide dispersion. If the
core refractive index is explicitly dependent on the wavelength,
$n(\lambda)$, photons of different wavelengths have different
propagation velocities along the same path, called material
dispersion. The broadening of a pulse travelling along a fibre is
given by $\Delta \tau = L_F/c_{\it core} \left( \lambda^2
d^2n/d\lambda^2 \right) \Delta\lambda/\lambda$, where $\Delta \lambda$
is the spectral width of the emission peaks of scintillating or
wavelength-shifting fibres~\cite{Ghatak1998}.  The full widths at half
maximum (FWHM) of the emission bands in most fibre polymers (e.g.\
polystyrene) is approximately $40-50$\,nm and the material dispersion
is of the order of a few ns$/$nm per kilometre fibre length and is
almost negligible for multimode fibres.

\begin{figure}[htbp]
  \begin{center}
      \epsfig{width= 0.5 \textwidth, file= 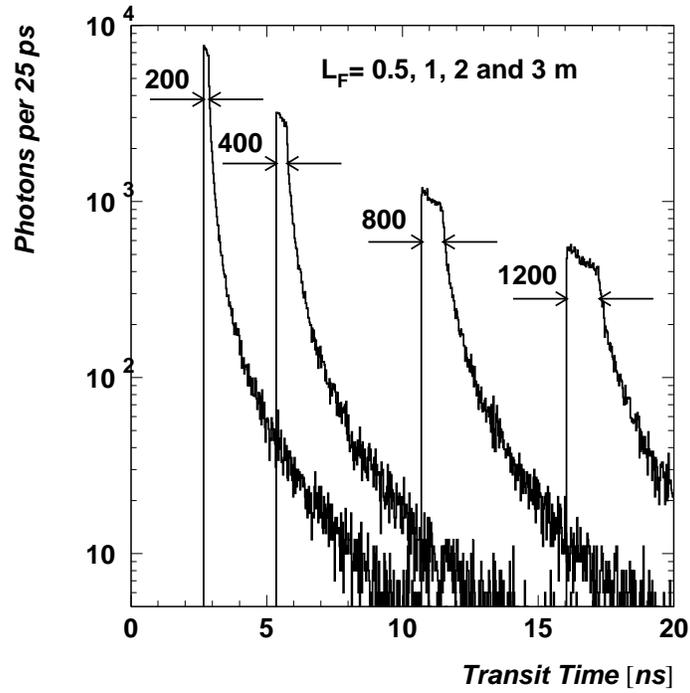}\\[5mm]
      \caption{ The distribution of the transit time in nanoseconds 
	for photons reaching the fibre exit end. For fibre lengths
	$L_F=$ 0.5, 1, 2 and 3\,m the pulse dispersion (FWHM) of the 
	transit time distribution is 200, 400, 800, and 1200\,ps, 
	respectively.}
      \label{fig:timing}
  \end{center}
\end{figure}

Because of the linear dependence of the transit time on the path
length the transit time is simply given by $\tau= L_F\,
P(\theta)/c_{\it core}$, where $c_{\it core}$ is the speed of light in
the fibre core and $L_F$ is the total axial length of the fibre. The
simulation results on the transit time are shown in
figure~\ref{fig:timing}. The FWHM of the pulses in the time spectrum
are presented for four different fibre lengths. The resulting
dispersion has to be compared with the time dispersion in the
meridional approximation which is simply the difference between the
shortest transit time $\tau_S = \tau(\theta=0)$ and the longest
transit time $\tau_L = \tau(\theta=\theta_{\it crit})$: $\Delta \tau
\equiv \tau_L - \tau_S = L_F/c_{\it core}\ (\sec{\theta_{\it
crit}}-1)$. The dispersion evaluates for ``standard'' fibres to
197\,ps for 0.5\,m, 393\,ps for 1\,m, 787\,ps for 2\,m and 1181\,ps
for 3\,m. Those numbers are in good agreement with the simulation,
although there are tails associated to the propagation of skew
rays. Using average attenuation parameters as discussed in section~4
the fraction of photons arriving later than $\tau_L$ decreases from
37.9\% for a 0.5\,m fibre to 32\% for a 3\,m fibre due to the stronger
attenuation of the skew rays in the tail. Because of the inter-modal
dispersion the pulse broadening in multimode fibres is quite
significant and for longer fibres ($L_F \ge$ 1\,m) the light
dispersion due to path length variations, $\sigma_{path}$, dominates
the timing resolution.

In the meridional approximation and assuming an isotropic emission,
$dN/d\Omega = 1/(4\pi)$, the probability of a photoelectron to get
emitted before a time $t$ can be derived. The normalised transit time
distribution in this approximation is $f(t) = \tau_S
\tau_L/(\Delta\tau\, t^2)$, leading to the probability
\begin{equation} 
  p(t) = \int_{\tau_L}^{t} f(t^\prime)\ dt^\prime = \frac{\tau_S
    \tau_L}{\Delta\tau} \left( \frac{1}{\tau_L} - \frac{1}{t} \right)\ .
\end{equation}
Then, the average transit times $\langle t \rangle = \int_0^1 t(p)\,
dp$ and $\langle t^2 \rangle = \int_0^1 t^2(p)\, dp$ can be
calculated, resulting in a time dispersion due to path length
differences:
\begin{equation}
  \sigma_{path}^2  =  \langle t^2 \rangle - \langle t \rangle^2
   =  \tau_S \tau_L - \left( \frac{\tau_S \tau_L}{\Delta \tau}
  \right)^2 \ln{\frac{\tau_L}{\tau_S}}\ .
\end{equation}
Recalling that $\tau_S \propto L_F$ and $\tau_L \propto L_F$ will
reproduce the linear dependence of the time dispersion on the distance
between the emission point and the photon sensor measured along the
fibre axis.

In order to reduce noise from dark counts the threshold of a
discriminator used with a time-to-digital converter is usually larger
than the minimum voltage corresponding to one photoelectron. Assuming
that the time measurement requires a fixed number of photoelectrons,
$j$, and $N$ photoelectrons are produced in a given event, then simple
probabilistic considerations provide the time distribution of
photoelectron $j$:
\begin{equation}
   P(p) = \frac{N!}{(N-j)!\, (j-1)!} (1-p(t))^{N-j}\, p(t)^{j-1}\ .
\end{equation}
Calculating the time dispersion in the same way as for the single
photoelectron case involves Gauss's hyper geometric function, $_2F_1$,
which arises frequently in physical problems,
\begin{equation}
  _2F_1(1, j, N+1, \Delta \tau / \tau_L) =
    \frac{\Gamma(N+1)}{\Gamma(j)\, \Gamma(N+1-j)} \int_0^1
    \frac{\tau_S}{1 - p \Delta \tau / \tau_L} (1-p)^{N-j}\,
    p^{j-1}\, dp\ ,
\end{equation}
where $\Gamma$ is Euler's Gamma function. Obviously, the integral is
dependent on $j$, and so the statistical behaviour of the timing
resolution is a function of the threshold. A parameterisation of the
integral with $\sigma_{path} \propto L_F/\sqrt{\bar{N}}$ is common. For
a large number of photoelectrons the approximation $\sigma_{path}
\propto L_F/\bar{N}$ is better suited.

The maximum axial angle allowed by the critical angle condition is
smaller in fibres than in bulk scintillators by
\begin{equation}
  \theta_{\it crit}^{bulk}/\theta_{\it crit}^{fibre} \approx
  \frac{\cos^{-1}1/n_{\it core}}{\cos^{-1}n_{\it clad}/n_{\it core}}
  \approx \sqrt{\frac{n_{\it core} - 1}{n_{\it core} - n_{\it clad}}}\ ,
\end{equation}
which equals $\approx 2.3$ for ``standard'' fibres. Thus, the time
dispersion due to path length variation is significantly reduced in
fibre bundles. A pioneering work by Kuhlen and
co-workers~\cite{Kuhlen1991} has shown that the timing resolution
achieved with a scintillating fibre detector is about a factor two
better than the timing resolution achieved with a geometrically
identical bulk scintillator.

\section{Conclusions}

Since the first demonstration of a scintillating fibre detector in the
early 1980s, particle detection and read-out techniques using active
fibres have become mainstream. Developments were mainly based on
step-index fibres with polystyrene cores.  The plastic scintillators
in active fibres are composites of more than one type of fluorescent
molecules containing aromatic rings. The energy transfer mechanisms in
these binary or even tertiary mixtures are complex, involving
radiationless internal conversions and overlapping absorption and
emission bands.

Nowadays, active optical fibres are an integral part of hadron
calorimetry. In addition, tracking detectors comprising thousands of
scintillating fibres are frequently built into large detector
systems. Wavelength-shifting fibres are found in several modern
particle physics experiments around the world. They allow for a
read-out with a very high level of hermeticity. Recently, the ongoing
interest concentrated on the development of low-mass particle
detectors with high timing and spatial resolutions.

Because of the relatively low quantum efficiency of the scintillator
and the low trapping efficiencies in fibres, the light yield at the
end of a long fibre is typically small. Thus, a detailed understanding
of the photon propagation characteristics is almost inevitable. In
this paper, the propagation of photons in straight and curved optical
fibres has been reviewed. The geometrical conditions have been
illustrated to explain the quantitative difference between meridional
and skew rays. The overall transmission through a fibre cannot be
described by a simple exponential function of propagation
distance. One contribution to this effect is the large spread in
optical path lengths between the most meridional and most skew rays.

Since light pulses are often near the noise level of the measurement
system, the light yield can get critically low for sharply curved
fibres, e.g.\ as used in calorimeters with a large hermeticity. A
Monte Carlo programme has been used to evaluate the loss of photons
propagating in fibres curved in a circular path in one plane. The
results show that the loss of photons due to the curvature of the
fibre is a simple function of radius of curvature to fibre radius
ratio and the transmission is $T > 90\%$ if the ratio is $\rho/R_{\it
curv} > 65$. The simulations also show that for larger ratios this
loss takes place in a transition region during which a new
distribution of photon angles is established. Photons which survive
the transition region propagate without further losses.

It is long known that long fibres have better timing resolutions than
bulk scintillation counters of the same overall dimensions. Fast
photon sensors with transit time spreads $\sigma_{TTS} \approx
0.1-0.5\,$ns in conjunction with plastic scintillators with time
constants $\sigma_{\it sci} < 2\,$ns allow the measurement of time
differences on the level of $\Delta\tau \simeq 100\,$ps.  Several
effects are responsible for the signal time spread in active
fibres. There are variations in the response time of the scintillator,
which include time variations in the energy transfer and the finite
decay time of the fluorescent dyes. Effects due to the light detection
include variations in the transit time from the photocathode to the
first dynode of photomultiplier tubes. Another group of limitations is
due to the electronics circuits used to process the signal. One of the
main contributions to the finite timing resolution comes from the time
dispersion due to path length variation in the fibre. The simulation
has been used to investigate the dispersion of transit times of
photons propagating in straight fibres. For fibre lengths between 0.5
and 3\,m approximately two thirds of the photons arrive within the
spread of transit times which would be expected from the use of the
simple meridional approximation and the refractive index of a
``standard'' fibre core. The remainder of the photons arrive in a tail
at later times due to their helical paths in the fibre. The fraction
of photons in the tail of the distribution decreases only slowly with
increasing fibre length and will depend on the attenuation parameters
of the fibre.

In conclusion, the timing properties of today's commercially available
fibres and photomultiplier tubes allow resolutions in the
sub-nanosecond region and the small detector dimensions make them very
attractive for experiments at present and future high luminosity
accelerators.

\acknowledgments I wish to express my thanks to J.H.~Cobb who
initiated my works on optical fibres during my years at Oxford
University.

`
\newpage


\end{document}